\documentclass{vldb}

\usepackage{color}
\usepackage[vlined, linesnumbered]{algorithm2e}
\usepackage{amsmath}
\usepackage{stmaryrd}
\usepackage{times}
\usepackage[normalem]{ulem}
\usepackage{titlesec}
\titlespacing{\section}{0pt}{*2}{*1}
\titlespacing{\subsection}{0pt}{*1.5}{*.5}
\usepackage{enumitem}

\usepackage[
pageanchor=true,
plainpages=false,
pdfpagelabels, 
bookmarks,
bookmarksnumbered,
pdfborder=0 0 0,
pdfpagemode=UseOutlines]{hyperref}

\newcommand{\csbbox}{\vrule height7pt width4pt depth1pt}

\newtheorem{thm}{Theorem}[section]
\newtheorem{lem}{Lemma}[section]

\newtheorem{CSEXAMPLE}{Example}[section]
\newenvironment{csexample}{\begin{CSEXAMPLE} \rm}%
                            {\end{CSEXAMPLE}}
\newcommand{\csxam}{\begin{csexample}}
\newcommand{\csexam}{\csbbox\end{csexample}}
\numberwithin{equation}{section}

\newcommand{\csprf}{

{\it Proof: }}

\newcommand{\csprfsketch}{

{{\it Proof sketch:} }}
\newcommand{\cseprf}{\csbbox\vpar}

\newcounter{ccc}

\newcommand\vpar{{\vspace{0.5em}}}
\newcommand\stitle[1]{{\vpar\noindent{\bf #1.}}}

\newcommand\queryplan[3][@R=8pt @C=2pt]{{\small \begin{minipage}[t]{#2}\xymatrix
    #1 {#3}\end{minipage}}}
\newcommand\querytree\queryplan

\newcommand\eat[1]{}

\newenvironment{smallpar}{

\small\sspace}{}

\newcommand{\otspace}{\vspace*{-1.2em}}
\newcommand{\ospace}{\vspace*{-1em}}
\newcommand{\espace}{\vspace*{-.8em}}
\newcommand{\sspace}{\vspace*{-.6em}}
\newcommand{\fspace}{\vspace*{-.4em}}
\newcommand{\tspace}{\vspace*{-.2em}}

\newenvironment{myalgorithmic}{%
\algorithm
\small
}{%
\endalgorithm
}
\newcommand{\FunctionTitle}[1]{
  \everypar={\relax}
  \setcounter{AlgoLine}{0} 
  \textbf{Function }#1 \\
}
\newcommand{\ProcedureTitle}[1]{
  \everypar={\relax}
  \setcounter{AlgoLine}{0} 
  \textbf{Procedure }#1 \\
}
\newcommand{\FuncSep}{
  \BlankLine
  \BlankLine
  \BlankLine
}
\SetKwInput{KwInit}{Initialization}
\SetKwInput{KwKnobs}{Knobs}

\SetCommentSty{smallCommentText}

\newcommand{\captioncell}[3]{\parbox[t][][t]{#1}{\centering \caption{#2}\label{#3}}}

\newenvironment{itemize*}{\begin{itemize}[leftmargin=1.5em, itemsep=2pt, topsep=3pt]}{\end{itemize}}

\newenvironment{description*}{\begin{description}[leftmargin=1.5em, itemsep=2pt, topsep=3pt]}{\end{description}}

\newenvironment{enumerate*}{\begin{enumerate}[leftmargin=1.5em, itemsep=2pt, topsep=3pt]}{\end{enumerate}}

\newif\ifshowadd
\newif\ifshowrem

\newcommand\edit[2]{%
	\ifshowadd%
		\textcolor{blue}{#1}%
	\else%
		{#1}%
	\fi%
    \ifshowrem%
        { \textrm{\sout{#2}}}%
    \fi%
}

\newcommand\rem[1]{%
	\ifshowrem%
		\textrm{\sout{#1}}%
	\fi%
}

\newcommand\add[1]{%
	\ifshowadd%
		\textcolor{blue}{#1}%
	\else%
		{#1}%
	\fi%
}

\newcommand{\calU}{{\mathcal U}}
\newcommand{\calC}{{\mathcal C}}
\newcommand{\calD}{{\mathcal D}}
\newcommand{\calM}{{\mathcal M}}
\newcommand{\calI}{{\mathcal I}}
\newcommand{\calJ}{{\mathcal J}}

\newcommand{\boldp}{\mbox{\bf p}}
\newcommand{\boldw}{\mbox{\bf w}}
\newcommand{\boldx}{\mbox{\bf x}}

\newcommand{\algo}{\mbox{\sc wfit}}
\newcommand{\WFA}{\mbox{\sc wfa}}
\newcommand{\wfaplus}{\WFA^{\scriptscriptstyle \! +}}
\newcommand{\OPT}{\mbox{\sc opt}}
\newcommand{\BASE}{\mbox{\sc base}}
\newcommand{\BC}{\mbox{\sc BC}}

\newcommand{\GOOD}{\scriptscriptstyle\rm GOOD}
\newcommand{\BAD}{\scriptscriptstyle\rm BAD}

\newcommand{\cost}{\mathit{cost}}
\newcommand{\benefit}{\mathit{benefit}}
\newcommand{\work}{w}
\newcommand{\doi}{\mathit{doi}}
\newcommand{\trans}{\delta}
\newcommand{\totwork}{\mathit{totWork}}

\newcommand{\score}{\mathit{score}}
\newcommand{\pdoi}{\doi^*}
\newcommand{\pben}{\benefit^*}

\newcommand{\analyzeQuery}{\mathit{analyzeQuery}}
\newcommand{\genRecommendation}{\mathit{recommend}}
\newcommand{\feedback}{\mathit{feedback}}
\newcommand{\repartition}{\mathit{repartition}}
\newcommand{\chooseCands}{\mathit{chooseCands}}
\newcommand{\extractIndices}{\mathit{extractIndices}}
\newcommand{\topIndices}{\mathit{topIndices}}
\newcommand{\choosePartition}{\mathit{choosePartition}}

\newcommand{\histSize}{\mathit{histSize}}
\newcommand{\randCnt}{\mathrm{RAND\_CNT}}
\newcommand{\idxCnt}{\mathit{idxCnt}}
\newcommand{\stateCnt}{\mathit{stateCnt}}
\newcommand{\intStats}{\mathit{intStats}}
\newcommand{\idxStats}{\mathit{idxStats}}
\newcommand{\loss}{\mathit{loss}}

\newcommand{\Scur}{\mathit{currRec}}
\newcommand{\Tcur}{\mathit{newRec}}

\newcommand{\cmax}{{c_\mathrm{max}}}
\newcommand{\tra}{\hspace{-0.6ex}\shortrightarrow\hspace{-0.6ex}}

\showaddfalse
\showremfalse

\begin{document}
	
\title{Semi-Automatic Index Tuning: Keeping DBAs in the Loop}

\numberofauthors{2}
\author{
  \alignauthor 	
	Karl Schnaitter \\ 
		\affaddr{Aster Data} \\ 
		\email{karl.schnaitter@asterdata.com}
  \alignauthor 	
	Neoklis Polyzotis \\ 
		\affaddr{UC Santa Cruz} \\ 
		\email{alkis@ucsc.edu}
}
    
  \maketitle
    
  \begin{abstract}
    To obtain good system performance, a DBA must choose a set of indices that is
appropriate for the workload. The system can aid in this challenging task by
providing recommendations for the index configuration. 
We propose a new index recommendation technique, termed semi-automatic tuning,
that keeps the DBA ``in the loop'' by generating recommendations that use
feedback about the DBA's preferences.
The technique also works online, which avoids the limitations of
commercial tools that require the workload to be known in advance. 
The foundation of our
approach is the Work Function Algorithm, which can solve a wide variety of
online optimization problems with strong competitive guarantees. We present an
experimental analysis that validates the benefits of semi-automatic tuning in a
wide variety of conditions.

  \end{abstract}
  
  \section{Introduction} 
\label{sec:introduction}

Index tuning, i.e., selecting indices that are appropriate for the
workload, is a crucial task for database administrators (DBAs).
However, selecting the right indices is a very difficult optimization problem: there exists a very large number of candidate indices for a given schema, indices may benefit some parts of the workload and also incur maintenance overhead when the data is updated, and the benefit or update cost of an index may depend on the existence of other indices. Due to this complexity, an administrator often resorts to
automated tools that can recommend possible index configurations after
performing some type of workload analysis.

In this paper, we introduce a novel paradigm for index tuning tools that we term \emph{semi-automatic index tuning}. 
A semi-automatic index tuning tool generates index recommendations by analyzing the workload
online, i.e., in parallel with query processing, which allows the
recommendations to adapt to shifts in the running workload. The DBA may request a recommendation at any time and is responsible for selecting the indices to create or drop. The most important and novel feature of semi-automatic tuning is that the DBA can provide feedback on the recommendation, which is taken into account for subsequent recommendations. In this fashion, the DBA can refine the automated recommendations by passing indirect domain knowledge to the tuning algorithm. Overall, the semi-automatic paradigm offers a unique combination of very desirable features: the tuner analyzes the running workload online and thus \edit{relieves}{alleviates} the DBA from the difficult task of selecting a representative workload; the DBA retains total control over the performance-critical decisions to create or drop indices; and, the feedback mechanism couples human expertise with the computational power of an automated tuner to enable an iterative approach to index tuning.

We illustrate the main features of semi-automatic tuning with a simple example. Suppose that the semi-automatic tuner recommends to materialize
three indices, denoted $a$, $b$, and $c$.
The DBA may materialize $a$, knowing that it has 
negligible overhead for the current workload. We interpret this as 
implicit positive feedback for $a$. The DBA might also provide explicit 
negative feedback on $c$ because past experience has shown that it
interacts poorly with the locking subsystem. In addition, the DBA may provide
positive feedback for another index $d$ that can benefit the same queries as
$c$ without the performance problems. Based on this feedback,
the tuning method can bias its recommendations in favor of indices $a,d$
and against index $c$. For instance, a subsequent recommendation could be
$\{a,d,e\}$, where $e$ is an index that performs well with $d$. At the
same time, the tuning method may eventually override the DBA's feedback
and recommend dropping some of these indices if the workload provides evidence that they do not perform well.

\stitle{Previous Work}
Existing approaches to index selection fall in two paradigms, namely offline and online.
Offline techniques~\cite{1142549,DBLP:journals/pvldb/BrunoC08}
generate a recommendation by analyzing a representative workload provided by
the DBA, and let the DBA make the final selection of indices. However, the DBA
is faced with the non-trivial task of selecting a good representative workload.
This task becomes even more challenging in dynamic environments (e.g., ad-hoc
data analytics) where workload patterns can evolve over time. 

Online techniques~\cite{bc:icde07,mwd:ssdbm09,slss:smdb07,Polyzotis:2007lr}
monitor the workload and automatically create or drop indices.
Online monitoring is essential to handle dynamic workloads, and
there is less of a burden on the DBA since a representative workload is not required.
On the other hand, the DBA is now completely out of
the picture. DBAs are typically very careful with changes to a running system,
so they are unlikely to favor completely automated methods. 

None of the existing index tuning techniques achieves the
same combination of features as semi-automatic tuning.
Semi-automatic tuning starts with the best features from the two paradigms
(online workload analysis with decisions delegated to the DBA) and augments
them with a novel feedback mechanism that enables the DBA to interactively
refine the recommendations. We note that interactive index tuning
has been explored in the literature~\cite{5447800}, but previous
studies have focused on offline workload analysis.
Our study is the first to propose an online feedback mechanism that
is tightly coupled with the index recommendation engine. 

A closer look at existing techniques also reveals that they cannot easily 
be modified to be semi-automatic. For instance, a naive
approach to semi-automatic tuning would simply execute an online tuning
algorithm in the background and generate recommendations based on the current
state of the algorithm, but this approach ignores the fact that the DBA may
select indices that contradict the recommendation.
A key challenge of semi-automatic tuning is to adapt the recommendations
in a flexible way that balances the influence of the workload and feedback
from the DBA.

\stitle{Our Contributions}
We propose the $\algo$ index-tuning algorithm that realizes the new paradigm
of semi-automatic tuning. $\algo$ uses a principled framework to
generate recommendations that take the workload and user feedback
into account.
We can summarize the technical contributions of this paper
as follows:

\vspace{0.4em}
\noindent
$\bullet$
We introduce the new paradigm of semi-automatic index tuning in
Section~\ref{sec:sat}.
We identify the relevant design choices, provide a formal problem statement, and outline the requirements for an
effective semi-automatic index advisor. 	

\vspace{0.4em}
\noindent$\bullet$
We show that recommendations can be generated in a principled manner by an
adaptation of the Work Function Algorithm~\cite{Borodin:1998ec} ($\WFA$)
from the study of metrical task systems (Section~\ref{sec:wfa:basic}).
We prove that $\WFA$ selects recommendations with a guaranteed bound
on worst-case performance, which allows the
DBA to put some faith in the recommended indices.
The proof
is interesting in the broader context of online optimization, since
the index tuning problem does not satisfy the assumptions of the
original Work Function Algorithm for metrical task systems.

\vspace{0.4em}
\noindent
$\bullet$
We develop the $\wfaplus$ algorithm (Section~\ref{sec:wfa:parts}) which uses a
divide-and-conquer strategy with several instances of $\WFA$ on
separate index sets.
We show that $\wfaplus$ leads to improved running time and better guarantees on
recommendation quality, compared to analyzing all indices with a single instance
of $\WFA$.
The guarantees of $\wfaplus$ are significantly stronger compared to previous
works for online database tuning~\cite{bc:icde07,mwd:ssdbm09}, and are thus
of interest beyond the scope of semi-automatic index selection.

\vspace{0.4em}
\noindent
$\bullet$
We introduce the $\algo$ index-tuning algorithm that provides an end-to-end
implementation of the semi-automatic paradigm (Section~\ref{sec:wfit}).
The approach builds upon the framework of $\wfaplus$,
and couples it with two additional components:
a principled feedback mechanism that is tightly integrated with
the logic of $\wfaplus$, and an online algorithm to extract candidate
indices from the workload.
%

\vspace{0.4em}
\noindent
$\bullet$~We evaluate $\algo$'s empirical performance using \add{a prototype implementation over IBM DB2} (Section~\ref{sec:experimental_study}). Our results with dynamic workloads demonstrate that $\algo$ generates online index recommendations of high quality, even when compared to the best indices that could be chosen with advance knowledge of the complete workload. We also show that $\algo$ can benefit from good feedback in order to improve further the quality of its recommendations, but is also able to recover gracefully from bad advice. 

\eat{ 

\begin{figure}
  \centering
  
  \includegraphics[scale=0.5]{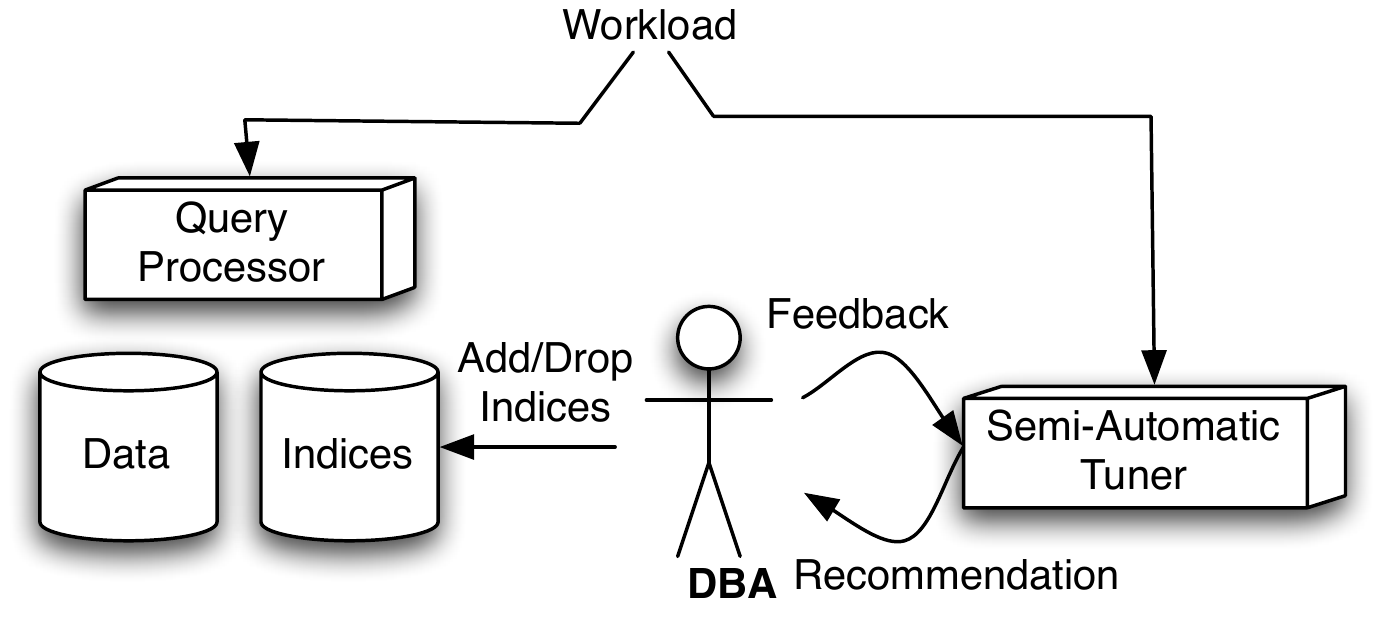}
  \fspace
  \caption{Overview of Semi-Automatic Tuning.}
  \otspace 
  \eat{
  \caption{The Tuner receives as input the stream of incoming queries and recommends on-demand a configuration for the materialized indices, based on a set of candidate indices and their interactions. The administrator may decide to modify the set of materialized indices, taking into account the recommendation and/or any other source of information, e.g., domain knowledge. These modifications are taken into account by the Tuner when generating a new recommendation. Thus, there is a direct feedback loop between the tuning process and the choices of the administrator.}}
  \label{fig:overview}
\end{figure}

%
%
%
Online
techniques~\cite{bc:icde07,mwd:ssdbm09,slss:smdb07,Polyzotis:2007lr}
monitor the workload and automatically create or drop indices.
This puts less of a burden on the DBA since a representative workload is not required,
but the drawback of online tuning is that the DBA is out of
the picture. DBAs are typically very careful with changes to a running system,
so they are unlikely to favor completely automated methods. 
One compromise between online and offline tuning is the Alerter proposed by 
Bruno and Chaudhuri~\cite{bc:vldb06}. This tool can be used to frequently 
analyze small batches of the workload and detect whether the existing index
configuration is suboptimal. If so, the tool will alert the DBA, who may then
perform a more thorough offline tuning analysis.

In this paper, we introduce a novel index tuning paradigm that we term
\emph{semi-automatic index tuning}. 
Figure~\ref{fig:overview} shows a high-level schematic of the approach. A
semi-automatic tuner generates recommendations by analyzing the workload
online, i.e., in parallel with query processing, which allows the
recommendations to adapt to shifts in the running workload. However, unlike
online tuning techniques, the recommendations are not implemented
instantly.  The DBA may request a recommendation at any time and
is responsible for selecting the indices to create or drop. 
This is vital in order for the DBA to retain control over the knobs that
affect system performance. The most important and novel feature of 
semi-automatic tuning is a mechanism to accept feedback
on the recommendation, which is taken into account for
subsequent recommendations. This feedback loop enables an iterative approach to
index tuning, where the DBA can refine the automated recommendations by passing
indirect domain knowledge to the tuning algorithm.

A naive approach to semi-automatic tuning would simply
execute an online tuning algorithm in the background,
and generate recommendations based on the current state of the algorithm.
This approach might be reasonable in the absence of user feedback or interventions.
The main challenge in semi-automatic tuning, however, is to adapt the recommendations in a flexible way
that balances the influence of the workload and feedback from the DBA.

We introduce the $\algo$ algorithm to realize the paradigm of semi-automatic
index selection. 
The logic that $\algo$ uses to analyze the workload and choose recommendations
is based on the well-known Work Function Algorithm~\cite{Borodin:1998ec}
from the study of metrical task systems.
This technique has strong performance guarantees, and we prove that the guarantees
extend to the problem of index selection. The proof is nontrivial, as index selection
cannot be modeled as a metrical task system.
The $\algo$ algorithm also builds upon the Work Function Algorithm with novel techniques
to reduce the computational complexity, improve the performance guarantees, incorporate
DBA feedback, and mine the workload for new candidate indices.
As a proof of concept, we have implemented a semi-automatic tuning prototype
based on $\algo$. In Section~\ref{sec:experimental_study} we present an experimental 
study that we performed using our prototype on workloads of
varying characteristics. The results validate the ability of $\algo$ to provide
meaningful recommendations that adapt to the workload and take the 
feedback of the DBA into account. 
} 


  \section{Preliminaries} 
\label{sec:preliminaries}

\stitle{General Concepts} 
We model the workload of a database
as a stream of queries and updates $Q$. 
We let $q_n$ denote the $n$-th statement
and $Q_N$ denote the prefix of length $N$.

Define $\calI$ as the set of secondary indices that may be created
on the database schema.
The physical database design comprises a subset of $\calI$ that may
change over time. 
Given a statement $q\!\in\!Q$ and set of indices $X \subseteq \calI$, we use $\cost(q,X)$ to
denote the cost of evaluating $q$ assuming that $X$ is the set of materialized
indices. This function is possible to evaluate through the what-if interface of
modern optimizers. Given disjoint sets $X,Y \subseteq \calI$, we define
$\benefit_q(Y,X) = \cost(q,X) - \cost(q,Y \cup X)$ as the difference in query
cost if $Y$ is materialized in addition to $X$. Note that $\benefit_q(Y,X)$ may be negative,
if $q$ is an update statement and $Y$ contains indices that need to be updated as a consequence of $q$.

Another source of cost comes from adding and removing materialized indices.
We let $\trans(X,Y)$ denote the cost to change the materialized set from $X$ to
$Y$. This comprises the cost to create the indices in $Y-X$ and to drop the
indices in $X-Y$. The $\trans$ function satisfies the triangle inequality:
$\trans(X,Y) \leq \trans(X,Z) + \trans(Z,Y)$.
However, $\trans$ is not a metric because indices are often far more expensive to 
create than to drop, and hence symmetry does not hold: $\trans(X,Y) \ne \trans(Y,X)$
for some $X,Y$.

\stitle{Index Interactions}
A key concern for index selection is the issue of
\emph{index interactions}. Two indices
$a$ and $b$ interact if the benefit of $a$
depends on the presence of $b$.
As a typical example, $a$ and $b$ can interact if they
are intersected in a physical plan, since the benefit of
each index may be boosted by the other. Note, however, that
indices can be used together in the same query plan without interacting.
This scenario commonly occurs when indices are used to handle selection
predicates on different tables.

We employ a formal model of index interactions that is based on our previous work on this topic~\cite{1687766}. Due to the complexity of index interactions, the model
restricts its scope to some subset $\calJ \subseteq \calI$ of interesting indices. (In our context, $\calJ$ is usually a set of indices that are relevant for the current workload.) The \emph{degree of interaction} between $a$ and $b$ with respect to a query $q$ is
\[
\doi_q(a,b) = \max_{X \subseteq \calJ} |\benefit_q(\{a\},X) - \benefit_q(\{a\},X \cup \{b\})|.
\]
\add{It is straightforward to verify the symmetry $\doi_q(a,b)=\doi_q(b,a)$ by expanding the expression of $\benefit_q$ in the metric definition.} Overall, this degree of interaction captures the amount that the benefits of
$a$ and $b$ affect each other. Given a workload $Q$, we say $a,b$ interact if $\exists q \in Q : \doi_q(a,b) > 0$,
and otherwise $a,b$ are independent.

Let $\{P_1,\dots,P_K\}$ denote a partition of indices in $\calJ$. 
Each $P_k$ is referred to as a \emph{part}. The
partition is called \emph{stable} if the cost function obeys the following identity for any $X \subseteq \calJ$:
\begin{eqnarray}
 \textstyle
 \cost(q,X) = \cost(q,\emptyset) - \sum_{k=1}^{K}\benefit_q(X \cap P_k,\emptyset).
 \label{eq:querycost}
\end{eqnarray}
Essentially, a stable partition decomposes the benefit of a large set $X$
into benefits of smaller sets $X \cap P_k$. The upshot for index tuning is that indices can be selected independently within each $P_k$, since indices from different parts have independent benefits. As shown in~\cite{1687766}, the stable partition
with the smallest parts is given by the connected components of the binary relation 
$\{ (a,b) ~|~ \mbox{$a,b$ interact} \}$. The same study also provides an efficient algorithm to compute the binary relation and hence the minimum stable partition.

In the worst case, the connected components can be quite large if there are
many complex index interactions.
In practice, the parts can be made smaller by ignoring weak interactions, i.e., index-pairs $(a,b)$ where $\doi_q(a,b)$ is small. Equation~(\ref{eq:querycost}) might not strictly hold in this case, but we can ensure that it provides a good
approximation of the true query cost (that is still useful for index tuning) as long as the partition
accounts for the most significant index interactions. We discuss this point in more detail in Section~\ref{sec:wfit}.


  \section{Semi-Automatic Index Tuning} 
\label{sec:sat}

At a high level, a semi-automatic tuning algorithm takes as input the current workload and feedback from the DBA, and computes a recommendation for the set of materialized indices. (Both inputs are continuous and revealed one ``element'' at a time.) The DBA may inspect the recommendation at any time, and is solely responsible for scheduling changes to the materialized set. The online analysis allows the algorithm to adapt its recommendations to changes in the workload or in the DBA's preferences. Moreover, the feedback mechanism enables the DBA to pass to the algorithm domain knowledge that is difficult to obtain automatically. We develop formal definitions for these notions and for the overall problem statement in the following subsection.

We note that our focus is on the core problem of generating index recommendations, which forms the basic component of any index advisor tool. An index advisor typically includes other peripheral components, such as a user interface to visually inspect the current recommendation~\cite{Hu:2008:QVQ:1454159.1454209,1687766} or methods to determine a materialization schedule for selected indices\cite{1687766}. These components are mostly orthogonal to the index-recommendation component and hence we can reuse existing implementations. Developing components that are specialized for semi-automatic index tuning may be an interesting direction for future work. 

\subsection{Problem Formulation}

\stitle{Feedback Model} We use a simple and intuitive feedback model that
allows the DBA to submit positive and negative votes according to current
preferences.
At a high level, a positive vote on index $a$ implies that we should favor
recommendations that contain $a$, until the workload provides sufficient
evidence that $a$ decreases performance. The converse interpretation is given
for a negative vote on $a$.
Our feedback model allows the DBA to cast several of these votes simultaneously.
Formally speaking, the DBA expresses new preferences 
by providing two disjoint sets of indices $F^+,F^- \subseteq \calI$,
where indices in $F^+$ receive positive votes and indices in
$F^-$ receive negative votes.

We say that the DBA provides \emph{explicit feedback} when they directly cast votes on indices. We also allow for \emph{implicit feedback} that can be derived from the manual changes that the DBA makes to the index configuration. More concretely, we can infer a positive vote when an index is created
and a negative vote when an index is dropped. The use of implicit feedback realizes an unobtrusive mechanism for automated tuning, where the tuning algorithm tailors its recommendations to the DBA's actions even if the DBA operates ``out-of-band'', i.e., without explicit communication with the tuning algorithm.

\stitle{Problem Formulation} A semi-automatic tuning algorithm receives as input
the workload stream $Q$ and a stream $V$ that represents the feedback provided
by the DBA. Stream $V$ has elements of the form $F=(F^+,F^-)$ per our feedback
model. Its contents are not synchronized with $Q$, since the DBA can provide
arbitrary feedback at any point in time.
We only assume that $Q$ and $V$ are ordered in time, and we may refer to
$Q \cup V$ as a totally ordered sequence.
The output of the algorithm is a stream of recommended index sets $S \subseteq \calI$,
generated after each query or feedback element in $Q \cup V$.
We focus on online algorithms, and hence the computation of $S$ can use
information solely from past queries and votes---the algorithm has absolutely
no information about the future. 

In order to complete the problem statement, we must tie the algorithm's output
to the feedback in $V$. Intuitively, we consider the DBA to be an expert and
hence the algorithm should trust the provided feedback. At the same time, the
algorithm should be able to recover from feedback that is not useful for the subsequent statements in the workload. We bridge these somewhat conflicting goals by
requiring each recommendation $S$ to be \emph{consistent} with recent feedback
in $V$.
To formally define consistency, let $F_c^+$ be the set of indices which have
received a vote after the most recent query, where the most recent vote was
positive.  Define $F_c^-$ analogously for negative votes.  The consistency
constraint requires $S$ to contain all indices in $F_c^+$
and no indices in $F_c^-$, i.e.,
$F_c^+ \subseteq S \land S \cap F_c^- = \emptyset$.

Consistency forces recommendations to agree with the DBA's cumulative feedback
so long as the algorithm has not analyzed a new query in the input. This
property is aligned with the assumption that the DBA is a trusted expert.
Moreover, consistency enables an intuitive interface in the case of implicit
feedback that is derived from the DBA's actions:
without the consistency constraint, it would be possible for the DBA to create
an index $a$ and immediately receive a recommendation to drop $a$
(an inconsistent recommendation) even though the workload has not changed.

\edit{At the same time, our definition implies that $F_c^+=F_c^-=\emptyset$ when a new query arrives.
This says that votes can only force changes to the recommended configuration
until the next query is processed, at which time the algorithm is given the
\emph{option} to override the DBA's previous feedback.
Of course, the algorithm needs to analyze the workload carefully before taking this option,
and determine whether the recent queries provide enough evidence to override past feedback.
Otherwise, it could appear to the DBA that the system is ignoring the feedback
and changing its recommendation without proper justification.
Too many changes to the recommendation can also hurt the theoretical
performance of an algorithm, as we describe later.
}
{The arrival of a new query nullifies the accumulated feedback ($F_c^+=F_c^-=\emptyset$ immediately after a query), and hence any recommendation satisfies the consistency constraint. Essentially, the algorithm is allowed to generate a recommendation that overrides the DBA's preferences if these are not useful for the subsequent statements in the workload. The semi-automatic tuning algorithm is given the freedom to determine whether the recent workload provides enough evidence to override past feedback.
}

\vpar
\textbf{The Semi-Automatic Tuning Problem:}
\emph{Given a workload $Q$ and a feedback stream $V$ of pairs
$(F^+,F^-)$,
generate a recommended index set $S \subseteq \calI$
after each element in $Q \cup V$ such that $S$
obeys the consistency constraint.}%
\vpar

Note that user-specified storage constraints are not part of the problem statement.
Although storage can be a concern in practice, the recommendation size
is unconstrained because it is difficult
to answer the question ``How much disk space is enough?''
before seeing the size of recommended indices.
Instead, we allow the DBA to control disk usage when selecting
indices from the recommendation.%
\footnote{Previous work~\cite{Hu:2008:QVQ:1454159.1454209,1687766} and
commercial systems provide tools to inspect index configurations,
which may be adapted to our setting.}
To validate our choice, we conducted a small survey among DBAs of real-world
installations. The DBAs were asked whether they would prefer to specify a space
budget for materialized indices, or to hand-pick indices from a recommendation
of arbitrary size. The answers were overwhelmingly in favor of the second
option.  One characteristic response said ``Prefer hand-pick from DBA
perspective, as storage is not so expensive as compared to overall objective of
building a highly scalable system.''
This does not imply we should recommend all
possible indices. On the contrary, as we see below, the recommendation must
account for the overhead of materializing and maintaining the indices it
recommends.

\stitle{Performance Metrics} Intuitively, a good semi-automatic tuning algorithm should 
recommend indices that minimize the overall
work done by the system, including the cost to process the workload as
well as the cost to implement changes to the materialized indices. The first component is typical for index tuning problems and it reflects the quality of the recommendations. The second component stems from the online nature of the problem: the recommendations apply to the running state of the system, and it is clearly desirable to change the materialized set at a low cost. Low materialization cost is important even if new indices are built during a maintenance period, since these periods have limited duration and typically involve several other maintenance tasks (e.g., generation of usage reports, or backups). 

Formally, let $A$ be a semi-automatic tuning algorithm, and define $S_n$ as the recommendation that $A$ generates after analyzing $q_n$ and all feedback up to $q_{n+1}$. 
Also denote the initial set of indices as $S_0$. We define the following \emph{total work} metric that
captures the performance of $A$'s recommendations:
\[
\totwork(A,Q_N,V)  =  
  \hspace{-.4em}
  \sum\limits_{1 \le n \le N} {\hspace{-.7em}
      \cost(q_n,S_n) + \trans(S_{n-1},S_n)}
\]
The value of $\totwork(A,Q_N,V)$ models the performance of a system where each
recommendation $S_n$ is adopted by the DBA for the processing of query $q_n$. This  convention follows common practice in the field of online algorithms~\cite{Borodin:1998ec} and is convenient for the theoretical analysis that we present later. In addition, this model captures the effect of the feedback in $V$, as each $S_n$ is required to be consistent (see above). Overall, total work forms an intuitive objective function, as it captures the primary sources of cost, while incorporating the effect of feedback on the choices of the algorithm. The adoption of this metric does not change the application of semi-automatic tuning in practice: the tuning algorithm will still generate a recommendation after each element in $Q \cup V$, and the DBA will be responsible for any changes to the materialized set.

\edit{
It is clearly impossible for an online algorithm $A$ to yield the optimal total work for all values of $Q_N$ and $V$. Consequently, we adopt the common practice of \emph{competitive analysis}: we measure the effectiveness of $A$ by comparing it against an idealized \emph{offline} algorithm $\OPT$ that has advance knowledge of $Q_N$ and $V$ and can thus generate optimal recommendations. Specifically, we say that $A$ has \emph{competitive ratio} $c$ if $\totwork(A,Q_N,V) \le c \cdot \totwork(\OPT,Q_N,V)+\alpha$ 
for any $Q_N$ and $V$, where $\alpha$ is constant with respect to $Q_N$ and $V$, and $A$ and $\OPT$ choose recommendations from the same finite set of configurations. The competitive ratio $c$ captures the performance of $A$ compared to the optimal recommendations in the \emph{worst case}, i.e., under some adversarial input $Q_N$ and $V$. In this work, we assume that $V=\emptyset$ for the purpose of competitive analysis, since $V$ comes from a trusted expert and hence the notion of adversarial feedback is unclear in practice. Our theoretical results demonstrate that the derivation of $c$ remains non trivial even under this assumption. Applying competitive analysis to the general case of $V \ne \emptyset$ is a challenging problem that we leave for future work.
}
{
Following common practice from online optimization, we can measure the effectiveness of an online algorithm $A$ by comparing it against an idealized algorithm $\OPT$ that has advance knowledge of the workload and feedback, and gives recommendations
that minimize total work. In order to make a meaningful comparison, we assume that
all recommendations are chosen from a finite set of configurations.
When $Q$ and $V$ are unknown, we can resort to \emph{competitive analysis}
to compare $A$ and $\OPT$ with adversarial input.
It would not be meaningful to consider adversarial feedback in this
setting, since we assume that $V$ corresponds to the feedback of a trusted expert.
Hence, for the purpose of competitive analysis, we set $V=\emptyset$ and
say $A$ has \emph{competitive ratio} $c$ if
$\totwork(A,Q_N,\emptyset) \le c \cdot \totwork(\OPT,Q_N,\emptyset)+\alpha$ 
for any workload sequence $Q_N$, where $\alpha$ is constant
with respect to $Q_N$. The ratio $c \ge 1$
captures $A$'s performance \emph{in the worst case} compared to the optimal
recommendations.
}

\begin{figure}
  \centering
  
  \includegraphics[scale=0.315]{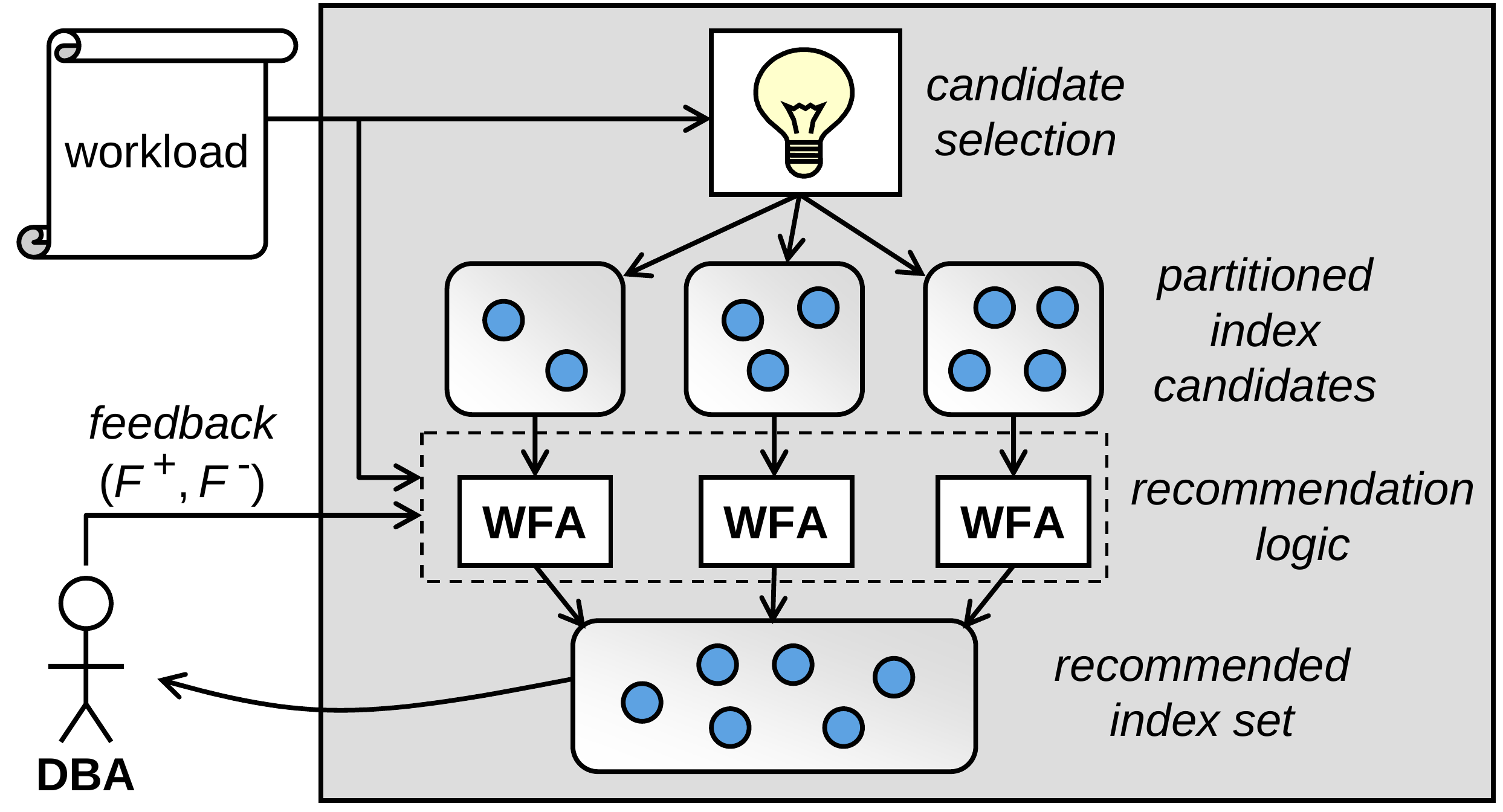}
  \caption{Components of the WFIT Algorithm.}
  \label{fig:design}
\end{figure}

\subsection{Overview of Our Solution}

The remainder of the paper describes the $\algo$ algorithm for semi-automatic index tuning. Figure~\ref{fig:design} illustrates $\algo$'s approach
to generating recommendations based on the workload and DBA feedback.
The approach starts with a \emph{candidate selection} component, which
generates indices that are relevant to the incoming queries. During candidate
selection, $\algo$ also analyzes the interactions between candidate indices
and uses these interactions to determine
a stable partition of the candidates~(see Section~\ref{sec:preliminaries}).
Then the output of candidate selection is a partitioned set of indices,
as shown in Figure~\ref{fig:design}.
Once these candidates are chosen, $\algo$
analyzes the benefit of the indices with respect to the workload
in order to generate the final recommendation.
The logic that $\algo$ uses to generate recommendations
is based on the Work Function Algorithm ($\WFA$)
of Borodin and El-Yaniv~\cite{Borodin:1998ec}. The original
version of $\WFA$ was proposed for metrical task systems~\cite{bls:jacm92} but
we extend its functionality to apply to semi-automatic index selection.
A separate instance of $\WFA$ analyzes each part of the candidate
set and only recommends indices within that part.
As we discuss later, this divide-and-conquer approach of $\algo$
improves the algorithm's performance and theoretical guarantees.
Finally, the DBA may request the current recommendation at any time
and provide feedback to $\algo$. The feedback is incorporated back
into each instance of $\WFA$ and considered for the next recommendation.

The following two sections present the full details of each component
of $\algo$ shown in Figure~\ref{fig:design}.
Section~\ref{sec:wfa} defines $\WFA$ and describes how $\algo$
leverages the array of $\WFA$ instances for its recommendation logic.
Section~\ref{sec:wfit} completes the picture, with the additional
mechanisms that $\algo$ uses to generate candidates and account
for DBA feedback.


  \section{A Work Function Algorithm \\ for Index Tuning}
\label{sec:wfa}

The index tuning problem closely follows the
study of task systems from online computation~\cite{bls:jacm92}.
This allows us to base our recommendation
algorithm on existing principled approaches. 
In particular, we apply the
Work Function Algorithm~\cite{Borodin:1998ec} ($\WFA$ for short), which is 
a powerful approach to task systems with an optimal competitive ratio.

\eat{
In Section~\ref{sec:wfa:basic}, we show how $\WFA$ can be tailored to the problem
of index tuning, which yields a very intelligent way to monitor the
workload and generate good recommendations.
The original competitive analysis of $\WFA$~\cite{Borodin:1998ec} 
does not apply in our setting, but we show how the proof may be extended to the
index recommendation problem.
We then describe an enhancement to $\WFA$ in Section~\ref{sec:wfa:parts},
which uses knowledge of index interactions to improve both the computational
complexity and the competitive ratio of $\WFA$. 
}

In order to fit the assumptions of $\WFA$, we do not consider the effect of feedback and we fix a set of candidate indices $\calC \subseteq \calI$ from which all recommendations will
be a drawn. In the next section, we will present the $\algo$ algorithm, which
builds on $\WFA$ with support for feedback and automatic maintenance of
candidate indices.

\subsection{Applying the Work Function Algorithm}
\label{sec:wfa:basic}

\begin{figure}[bp]
    \centering
    \includegraphics[scale=0.42]{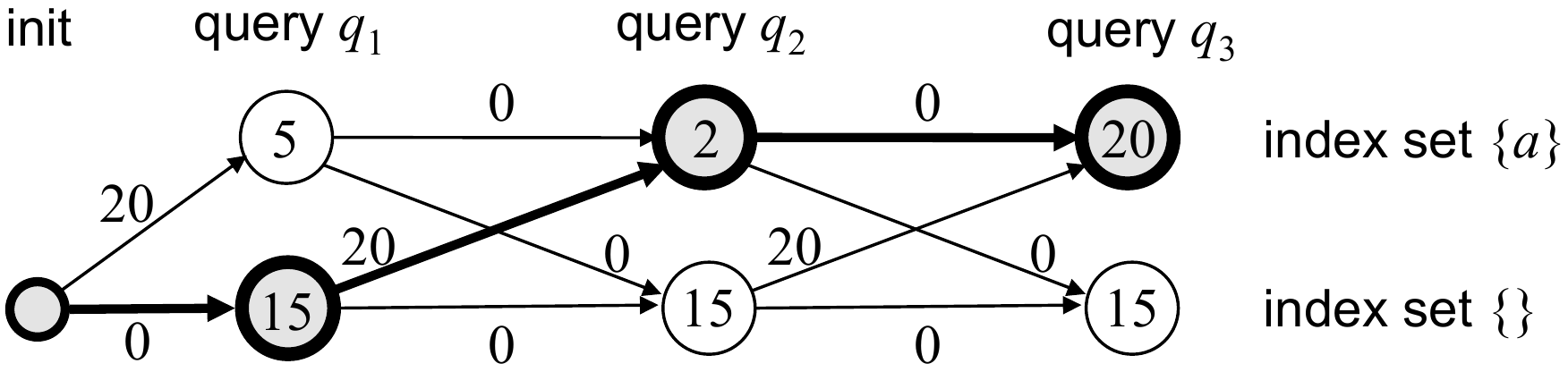}

    \smallskip
    \parbox{0.96\hsize}{
    \scriptsize
    This small graph visualizes total work for a workload of three queries
    $q_1,q_2,q_3$, where recommendations 
    are chosen between $\emptyset$ and $\{a\}$. 
    The index $a$ has cost 20 to create and cost 0 to drop. 
    The highlighted path in the graph corresponds to an algorithm that recommends
    $\emptyset$ for $q_1$ and $\{a\}$ for $q_2,q_3$. The combined cost of 
    edges and nodes in the path
    is $\trans(\emptyset,\emptyset) + \cost(q_1,\emptyset)+\trans(\emptyset,\{a\}) + \cost(q_2,\{a\}) + \trans(\{a\},\{a\}) + \cost(q_3,\{a\})=57$.}
    \caption{Index transition graph}
    \label{fig:transition_graph}
\end{figure}

We introduce the approach of $\WFA$ with a conceptual tool
that visualizes the index tuning problem in the form of a graph.
The graph has a source vertex $S_0$ to represent the initial state of
the system, as well as vertices $(q_n,X)$ for each statement $q_n$
and possible index configuration $X \subseteq \calC$. The graph has an edge from
$S_0$ to $(q_1,X)$ for each $X$, and edges from $(q_{n-1},X)$ to $(q_n,Y)$
for all $X,Y$ and $1 < n \leq N$. The weight of an edge is given by the 
transition cost between the corresponding index sets.
The nodes $(q,X)$ are also annotated with a weight of $\cost(q,X)$.
We call this the \emph{index transition graph}.
The key property of the graph is that the $\totwork$ metric is
equivalent to the sum of node and edge weights along the path
that follows the recommendations.
Figure~\ref{fig:transition_graph} illustrates this calculation
on a small sample graph.
A previous study~\cite{1142549} has used this graph formulation for
index tuning when the workload sequence is
known a priori. Here, we are dealing with an online setting where the workload
is observed one statement at a time.
\eat{and recommendations are required to be consistent with the DBA's feedback.}

The internal state of $\WFA$ records information about shortest paths in the
index transition graph, where the possible
index configurations comprise the subsets of the candidate set $\calC$.
More formally, after observing $n$ workload statements, the internal state of
$\WFA$ tracks a value denoted $\work_n(S)$ for each index set $S \subseteq \calC$,
as defined in the following recurrence:
\begin{eqnarray}
  \work_{n}(S)
  &=&
  \min_{X \subseteq \calC}\{ \work_{n-1}(X) + \cost(q_{n},X) + \trans(X,S)\}
  \ \ \ \ \ \ \ \ \ 
  \label{eq:work_function_recurrence}
  \\
  \work_0(S)
  &=&
  \trans(S_0,S)\nonumber  
\end{eqnarray}
We henceforth refer to $\work_n(S)$ as the work function value for $S$ after
$n$ statements.
As mentioned above, the work function can be interpreted in terms of paths in
the index transition graph.
In the case where $n$ is positive, $\work_n(S)$ represents the sum of
(i)~the cost of the shortest path from $S_0$ to some graph node $(q_n,X)$, and
(ii)~the transition cost from $X$ to $S$.
The actual value of $\work_n(S)$ uses the $X \subseteq \calC$ which minimizes
this cost. We can think of $w_0(S)$ in a similar way, where the ``path'' is
an empty path, starting and ending at $S_0$.
Then the definition $\work_0(S) = \trans(S_0,S)$ has a natural analogy
to the recursive case.

Note that the total work of the theoretically optimal recommendations
is equivalent to
$\totwork(Q_n,\OPT,\emptyset) = \min_{S \subseteq \calC}\{ \work_n(S) \}$. 
Hence, the intuition is that $\WFA$ can generate good
recommendations online by maintaining information about the possible paths
of optimal recommendations.

Figure~\ref{alg:WFA} shows the pseudocode for applying $\WFA$ to
index tuning.
All of the bookkeeping in $\WFA$ is based on the fixed set $\calC$
of candidate indices.
The algorithm records an array $\boldw$ that is indexed by the possible
configurations (subsets of $\calC$). After analyzing the $n$-th statement of
the workload, $\boldw[S]$ records the work function value $\work_n(S)$. The
internal state also includes a variable $\Scur$ to record the current
recommendation of the algorithm. 

The core of the algorithm is the $\analyzeQuery$ method. There are two stages to
the method. The first stage updates the array $\boldw$ using the recurrence
expression defined previously. The algorithm also creates an auxiliary
array~$\boldp$.
Each $\boldp[S]$ contains index sets $X$ such that a path from
$S_0$ to $(q_n,X)$ minimizes $\work_n(S)$.
\eat{ 
\begin{csexample}
We illustrate the first stage of $\analyzeQuery$ using the scenario in
Figure~\ref{fig:transition_graph}. Before the first
query is seen, note that the work function values are initialized as
as in the base case of~(\ref{eq:work_function_recurrence}). Assuming the
initial configuration is empty, we have
\begin{smallpar}\[
    \work_0(\emptyset) = \trans(\emptyset,\emptyset) = 0, \hspace{10pt} 
    \work_0(\{a\}) = \trans(\emptyset,\emptyset) = 20.
\]\end{smallpar}%
The entries of $\boldw$ are set to these two values.
After the first query, the work function for the empty configuration is
\begin{smallpar}
\begin{eqnarray*}
\work_1(\emptyset)
 &=&
 \min \{\work_0(X) + \cost(q_1,X) + \trans(X,\emptyset) ~|~ X = \emptyset, \{a\}\} \\
 &=&
 \min \{0 + 15 + 0, 20 + 5 + 0 \} \\
 &=& 15.
\end{eqnarray*}
\end{smallpar}%
The value of $\work_0(\emptyset)$ corresponds to the path along
the edge that stays in the empty configuration, and the node weight
accounts for the the entire cost of 15.
We follow similar steps to arrive at
\begin{smallpar}\[
\work_1(\{a\}) = 25.
\]\end{smallpar}%
\end{csexample}}
The second stage computes the next
recommendation to be stored in $\Scur$. $\WFA$ assigns a numerical score to
each configuration $S$ as $\score(S)=\boldw[S]+\trans(S,\Scur)$
and the next state must minimize this score.
To see the intuition of this criterion, 
consider a configuration $X$ with a higher score than $\Scur$,
meaning that $X$ cannot become the next
recommendation. Then
\[ \begin{array}{rl}
  & \score(\Scur) < \score(X) \\
  \Rightarrow &
  \work_n(\Scur)-\work_n(X) < \trans(X,\Scur).
  \end{array}\]
The left-hand side of the final inequality can be viewed as the benefit of
choosing a new recommendation $X$ over $\Scur$ in terms of the total work function,
whereas the right side represents the cost for $\WFA$ to 
``change its mind'' and transition from $X$ back to $\Scur$.
When the benefit is less than the transition cost, 
$\WFA$ will not choose $X$ over the current recommendation.  This 
cost-benefit analysis helps $\WFA$ make robust decisions 
(see Theorem~\ref{thm:wfa_competitive}). 

\begin{figure}
  \hspace{-1em}
  \begin{minipage}{2\hsize}
  \begin{myalgorithmic}
      \KwData{\hspace{-0.5ex}Set $\calC \subseteq \calI$ of candidate indices; 
       Array $\boldw$ of work function values;
       \\ \hspace{3ex}
       Configuration $\Scur$.}  
    \KwInit{Candidates $\calC$ and initial state $S_0 \subseteq \calC$ given as input;
       \\ \hspace{3ex}
       $\boldw[S]=\trans(S_0,S)$ for each $S \subseteq \calC$;
       $\Scur = S_0$.
       }
  \end{myalgorithmic}
  \end{minipage}
  \begin{myalgorithmic}
    \FuncSep

    \ProcedureTitle{\FuncSty{$\WFA.\analyzeQuery$}($q$)}
    \KwIn{The next statement $q$ in the workload}
    Initialize arrays $\boldw'$ and $\boldp$\;
    \ForEach{$S \subseteq \calC$} {
      $\boldw'[S] = \min_{X \subseteq \calC} \{ \boldw[X] + \cost(q,X) + \trans(X,S)\}$\;
      $\boldp[S] = \{X \subseteq \calC ~|~ \boldw'[S] = \boldw[X] + \cost(q,X) + \trans(X,S) \}$\;
    }
    Copy $\boldw'$ to $\boldw$\;
    \showlnlabel{ln:wfa_score}
    \lForEach{$S \subseteq \calC$} {$\score(S) \leftarrow \boldw[S] + \trans(S,\Scur)$}\;
    $\Scur \leftarrow \arg\min_{S \in \boldp[S]} \{\score(S)\}$\;

    \FuncSep

    \FunctionTitle{\FuncSty{$\WFA.\genRecommendation()$}}
    \setcounter{AlgoLine}{1} \showln 
    \Return{$\Scur$}\;

  \end{myalgorithmic}
  \caption{Pseudocode for $\WFA$.}
  \label{alg:WFA}
\end{figure}

The recommendation $S$ chosen by $\WFA$ 
must also appear in $\boldp[S]$. 
Recall that $\boldp[S]$ records states $X$ s.t.~there exists a path 
from $S_0$ to $(q_n,X)$ that minimizes $\work_n(S)$.
The condition specifies that $X = S$ for one such path,
and hence $\work_n(S) = \work_{n-1}(S) + \cost(q,S)$. 
An important result from Borodin et al.~(\cite{Borodin:1998ec}, Lemma~9.2)
shows that this condition is always satisfied by a state with minimum score.
In other words, the criterion $S \in \boldp[S]$ is merely a tie-breaker for
recommendations with the minimum score, to favor configurations whose work function
does not include a transition after the last query is processed. This is
crucial for the theoretical guarantees of $\WFA$ that we discuss
later.

\begin{csexample}
\label{ex:wfa}
The basic approach of $\WFA$ can be illustrated using the scenario in
Figure~\ref{fig:transition_graph}. The actual recommendations of
$\WFA$ will be the same as the highlighted nodes. Before the first
query is seen, the work function values are initialized as
\begin{smallpar}\[
    \work_0(\emptyset) = 0, \hspace{10pt} 
    \work_0(\{a\}) = 20
\]\end{smallpar}%
based on the transition cost from the initial configuration $S_0\equiv\emptyset$.
After the first query, the work function is updated 
using~\textup{(\ref{eq:work_function_recurrence}):}
\begin{smallpar}\[
    \work_1(\emptyset) = 15, \hspace{10pt}
    \work_1(\{a\}) = 25.
\]\end{smallpar}%
These values are based on the paths
$\emptyset \tra \emptyset$ and $\emptyset \tra \{a\}$
respectively.\footnote{For example~\ref{ex:wfa}, we abuse notation and
use index sets $X$ in place of the graph nodes $(q_n,X)$.}
The scores are the same as the respective work
function values \textup{(}$\trans(\emptyset,\emptyset)=\trans(\{a\},\emptyset)=0$ at
line~\ref{ln:wfa_score} of $\WFA.\analyzeQuery$\textup{)}, hence $\emptyset$ remains
as $\WFA$'s recommendation due to its lower score.
After $q_2$, the work function values are both
\begin{smallpar}\[
    \work_2(\emptyset) = \work_2(\{a\}) = 27.
\]\end{smallpar}%
Both values use the path $\emptyset \tra \{a\} \tra \{a\}$.
The calculation of $\work_2(\emptyset)$ also includes the transition
$\trans(\{a\},\emptyset)$, which has zero cost.
The corresponding scores are again equal to the work function,
but here the tie-breaker comes into play\textup{:}
$\{a\}$ is preferred because it is
used to evaluate $q_2$ in both
paths, hence $\WFA$ switches its recommendation to $\{a\}$.
Finally, after $q_3$, the work function values are
\begin{smallpar}\[
    \work_3(\emptyset) = 42, \hspace{10pt}
    \work_3(\{a\}) = 47.
\]\end{smallpar}%
based on paths
$\emptyset \tra \{a\} \tra \{a\} \tra \emptyset$ and
$\emptyset \tra \{a\} \tra \{a\} \tra \{a\}$ respectively.
The actual scores must also account for the current recommendation
$\{a\}$. Following line~\ref{ln:wfa_score} of $\WFA.\analyzeQuery$,
\begin{smallpar}\[
    \score(\emptyset) = 62, \hspace{10pt}
    \score(\{a\}) = 47.
\]\end{smallpar}%
The recommendation of $\WFA$ remains $\{a\}$, since it has a lower
score. This last query illustrates an interesting property of $\WFA$\textup{:}
although the most recent query has favored dropping $a$, the
recommendation does not change because the difference in work function 
values is too small to outweigh the cost to materialize $a$ again.~\csbbox
\end{csexample}

As a side note, observe that the computation of $\work_n(S)$ requires
computing $\cost(q,X)$ for multiple configurations $X$. 
This is feasible using the what-if optimizer of the database system. 
Moreover, recent studies~\cite{1325974,DBLP:conf/sigmod/BrunoN08}
have proposed techniques to speed up successive what-if optimizations
of a query. These techniques can readily be applied to make the
computation of $\work_n$ very efficient.

\stitle{WFA's Advantage: Competitive Analysis} $\WFA$ is a seemingly simple algorithm, but its key advantage is that we can prove strong guarantees on the performance of its recommendations. 

Borodin and El-Yaniv~\cite{Borodin:1998ec} showed that $\WFA$ has a competitive
ratio of $2 \sigma-1$ for any metrical task system with $\sigma$ possible configurations,
meaning that its worst-case performance
can be bounded. Moreover, $\WFA$ is an optimal online algorithm, as this is the best
competitive ratio that can be achieved. These are very powerful properties that
we would like to transfer to the problem of index recommendations. However, the
original analysis does not apply in our setting, since it requires $\trans$ to
be a metric, and our definition of $\trans$ is not symmetric.
One of the technical contributions of this paper is to show how to overcome the
fact that $\trans$ is not a metric, and extend the analysis to the problem of
index recommendations.

\begin{thm}
  \label{thm:wfa_competitive}
  The $\WFA$ algorithm, as shown in Figure~\ref{alg:WFA},
  has a competitive ratio of $2^{|\calC|+1}-1$. \emph{(Proof in the appendix)} 
\end{thm}

This theoretical guarantee bolsters our use of $\WFA$ to generate
recommendations. The competitive ratio ensures that the
recommendations do not have an arbitrary effect on performance in the worst case.
We show empirically in Section~\ref{sec:experimental_study} that the 
average-case performance of the recommendations can be close to optimal.
This behavior is appealing to DBAs, since they would not want to make changes
that can have unpredictably bad performance.

\subsection{Partitioning the Candidates}
\label{sec:wfa:parts}

In the study of general task systems, the competitive ratio of $\WFA$ is
theoretically optimal~\cite{bls:jacm92}.
However, the algorithm has some drawbacks for the index
recommendation problem, since it becomes infeasible to maintain
statistics for every subset of candidates in $\calC$ as the size of $\calC$ increases.
The competitive ratio $2^{|\calC|+1}-1$ also becomes nearly
meaningless for moderately large sets $\calC$.
Motivated by these observations, we present an enhanced algorithm $\wfaplus$, which
exploits knowledge of index interactions to reduce the computational complexity of $\WFA$,
while enabling stronger theoretical guarantees.

The strategy of $\wfaplus$ employs a stable partition $\{C_1,\dots,C_K\}$
of $\calC$, as defined in Section~\ref{sec:preliminaries}.
The stable partition guarantees that indices in $C_k$ do not
interact with indices in any other part $C_l \ne C_k$.
This is formalized by (\ref{eq:querycost}), which shows that
each part $C_i$ makes an independent contribution to the benefit. 
Moreover, it is
straightforward to show that $\trans(X,Y) = \sum_k \trans(X \cap C_k, Y\cap C_k)$, 
i.e., we can localize the transition cost within each subset $C_k$.
These observations allow $\wfaplus$ to decompose 
the objective function $\totwork$ into $K$ components, one for each $C_k$,
and then select indices within each subset using separate instances of $\WFA$.

We define $\wfaplus$ as follows.
The algorithm is initialized with a stable partition $\{C_1,\dots,C_K\}$ of $\calC$,
and initial configuration $S_0$. For $k=1,\dots,K$, $\wfaplus$ maintains
a separate instance of $\WFA$, denoted $\WFA^{(k)}$.
We initialize $\WFA^{(k)}$ with candidates
$C_k$ and initial configuration $S_0 \cap C_k$. The interface of
$\wfaplus$ follows $\WFA$:
\begin{itemize*}
\item
    $\wfaplus.\analyzeQuery(q)$ calls $\WFA^{(k)}.\analyzeQuery(q)$ for each $k=1,\dots,K$.
\item
    $\wfaplus.\genRecommendation()$ returns  
    $\bigcup_k \WFA^{(k)}.\genRecommendation()$.
\end{itemize*}

On the surface, $\wfaplus$ is merely a wrapper around multiple
instances of $\WFA$, but the partitioned approach of $\wfaplus$ 
provides several concrete advantages.
The division of indices into a stable partition
implies that $\wfaplus$ must maintain statistics on only
$\sum_k 2^{|C_k|}$ configurations, compared to
the $2^{|\calC|}$ states that would be required to monitor all the indices
in $\WFA$. This can simplify the bookkeeping
massively: a back-of-the-envelope calculation shows that if $\wfaplus$ is given
32 indices partitioned into subsets of size 4, then only 128 configurations need
to be tracked, whereas $\WFA$ would require more than four billion states.
We prove that this simplification is lossless, i.e., that 
$\wfaplus$ selects the same indices as $\WFA$.

\begin{thm} \label{thm:wfit_partition}
  If $\{C_1,\dots,C_K\}$ is a stable partition of $\calC$, 
  then $\wfaplus$ on $\{C_1,\dots,C_K\}$ will make the same recommendations
  as $\WFA$ on $\calC$. 
  \emph{(Proof in the appendix)}
\end{thm}

It immediately follows that $\wfaplus$ inherits the competitive ratio of $\WFA$.
However, the power of $\wfaplus$ is that 
it enables a much smaller competitive ratio
by taking advantage of the stable partition.
\begin{thm}
  \label{thm:wfit_competitive}
  $\wfaplus$ has a competitive ratio of $2^{\cmax+1}-1$, where $\cmax=\max_k\{|C_k|\}$. 
  \emph{(Proof in the appendix)}
\end{thm}

Hence the divide-and-conquer strategy of $\wfaplus$ is a win-win, as it improves the 
computational complexity of $\WFA$ as well as the guarantees on performance.
Observe that $\wfaplus$ matches the competitive ratio of 3 that the online tuning algorithm
of Bruno and Chaudhuri~\cite{bc:icde07} achieves for the special case $|\calC|=1$
(the competitive analysis in~\cite{bc:icde07} does not extend to
a more general case).
The competitive ratio is also superior to the ratio $\ge 8(2^{|\calC|}-1)$ 
for the OnlinePD algorithm of Malik et al.~\cite{mwd:ssdbm09}
for a related problem in online tuning.

  \section{The WFIT Algorithm}
\label{sec:wfit}

We introduced $\wfaplus$ in the previous section, as a solution to the 
index recommendation problem with strong theoretical guarantees.
The two limitations of $\wfaplus$ are (i) it does not accept feedback,
and (ii) it requires a fixed set of
candidate indices and stable partition.
In this section, we define the $\algo$ algorithm, which
extends $\wfaplus$ with mechanisms to incorporate feedback
and automatically maintain the candidate indices.

Figure~\ref{alg:wfit} shows the interface of $\algo$
in pseudocode. The methods $\analyzeQuery$ and
$\genRecommendation$ perform the same steps as the corresponding
methods of $\wfaplus$.
In $\analyzeQuery$,
$\algo$ takes additional steps to 
maintain the stable partition $\{C_1,\dots,C_K\}$.
This work is handled by two auxiliary methods:
$\chooseCands$ determines what the next partition should be, and
$\repartition$ reorganizes the data structures of $\algo$ for the 
new partition. Finally, $\algo$ adds a new method
$\feedback$, which incorporates explicit or implicit feedback
from the DBA.

In the next subsection, we discuss the $\feedback$ method. We then
provide the details of the $\chooseCands$ and $\repartition$ methods
used by $\analyzeQuery$.

\subsection{Incorporating Feedback}
\label{sub:feedback}

As discussed in Section~\ref{sec:sat}, the DBA provides feedback by casting  positive votes for indices
in some set $F^+$ and negative votes for a disjoint set $F^-$.
The votes may be cast at any point in time, and the sets $F^+,F^-$
may involve any index in $\calC$
(even indices that are not part of the current recommendation).
This mechanism is captured by a new method $\feedback(F^+,F^-)$.
The DBA can call $\feedback$ explicitly to express preferences about
the index configuration, and we also use $\feedback$
to account for the implicit feedback from manual changes to the
index configuration.

Recall from Section~\ref{sec:sat} that the recommendations
must be \emph{consistent} with recent feedback,
but should also be able to \emph{recover} from poor feedback.
Our approach to guaranteeing consistency is simple:
Assuming that $\Scur$ is the current recommendation, the new recommendation
becomes $\Scur - F^- \cup F^+$. 
Since $\algo$ forms its recommendation as
$\bigcup_k \Scur_k$, where $\Scur_k$ is the recommendation
from $\WFA$ running on part $C_k$, we need to modify each $\Scur_k$
accordingly. Concretely, the new recommendation for $C_k$ becomes
$\Scur_k - F^- \cup (F^+ \cap C_k)$. 

The recoverability property is trickier to implement properly. 
Our solution is to adjust the scores in order to appear as if
the workload (rather than the feedback) had led $\algo$ 
to recommend creating $F^+$ and dropping $F^-$.
With this approach, $\algo$ can naturally recover from bad feedback if
the future workload favors a different configuration.
To enforce the property in a principled manner, we
need to characterize the internal state of each instance of $\WFA$ after it generates a
recommendation. Recall that $\WFA$ selects its next recommendation as the
configuration that minimizes the $\score$ function. Let us assume that the
selected configuration is $Y$, which differs from the previous configuration by
adding indices $Y^+$ and dropping indices $Y^-$. If we
recompute $\score$ after $Y$ becomes the current recommendation, then we can
assert the following bound for each configuration $S$: 
\begin{eqnarray}
& & \hspace{-4em}
    \mbox{\small $\score(S) - \score(Y) \ge$} \nonumber \\
& & \hspace{1em} 
    \mbox{\small $\trans(S, S - Y^- \cup Y^+) + \trans(S - Y^- \cup Y^+, S)$}
\label{eq:sat:threshold}
\end{eqnarray}
Essentially, this quantity represents the
minimum threshold that $\score(S)$ must overcome in order to replace the recommendation $Y$.
Hence, in order for the internal state of $\WFA^{(k)}$ to be
consistent with switching to the new recommendation $\Scur_k$, we must ensure that
$\score(S) - \score(\Scur_k)$, or the equivalent expression
$\boldw^{(k)}[S] + \trans(S,\Scur_k) - \boldw^{(k)}[\Scur_k]$,
respects~(\ref{eq:sat:threshold}).
This can be achieved by increasing $\boldw^{(k)}[S]$ accordingly. 

\begin{figure}[t]
  \begin{myalgorithmic}
    \KwData{Current set $\calC$ of candidate indices;
       \\ \hspace{3ex}
       Stable partition $\{C_1,\dots,C_K\}$ of $\calC$;
       \\ \hspace{3ex}
       $\WFA$ instances $\WFA^{(1)},\dots,\WFA^{(K)}$;
    }
    \KwInit{
       Initial index set $S_0$ is provided as input;
       \\ \hspace{3ex}
       $\calC = S_0$, \add{$K=|S_0|$} and \edit{$C_i = \{a_i\}$ where $1 \le i \le |S_0|$ and  $a_1,\dots,a_{|S_0|}$ are the indexes in $S_0$ }{$ \{C_1,\dots,C_K\} = \{\{a\} ~|~ a \in S_0\}$};
       \\ \hspace{3ex}
       \begin{minipage}[t]{75mm}
       \For{$k\leftarrow$ $1$ \KwTo $K$}{
          $\WFA^{(k)} \leftarrow \mbox{
          \begin{minipage}[t]{50mm}
          instance of $\WFA$ with candidates $C_k$ \\
          and initial configuration $C_k \cap S_0$
          \end{minipage}}$
       }
       \end{minipage}
    }

    \FuncSep

    \ProcedureTitle{\FuncSty{$\algo.\analyzeQuery$}($q$)}%
    \KwIn{The next statement $q$ in the workload.}
      $\{D_1,\dots,D_M\} \leftarrow \chooseCands(q)$ \tcp*[l]{Figure~\ref{alg:auto_stable_part}}
      \If{$\{D_1,\dots,D_M\} \ne \{C_1,\dots,C_K\}$}
      {
		\tcp{Replace $\{C_1,\dots,C_K\}$ with $\{D_1,\dots,D_M\}$.}
        $\repartition(\{D_1,\dots,D_M\})$
        \tcp*[l]{Figure~\ref{alg:repartition}}
      }
      \lFor{$k\leftarrow$ $1$ \KwTo $K$}{$\WFA^{(k)}.\analyzeQuery(q)$\;}

    \FuncSep

    \FunctionTitle{\FuncSty{$\algo.\genRecommendation()$}}
    \setcounter{AlgoLine}{1} \showln 
    \Return{$\bigcup_k \WFA^{(k)}.\genRecommendation()$}\;

  \FuncSep

    \ProcedureTitle{\FuncSty{$\algo.\feedback$}($F^+,F^-$)}
    \KwIn{Index sets $F^+,F^- \subseteq \calC$ with positive/negative votes.}
    \For{$ k \leftarrow 1$ \KwTo $K$} {
      Let $\boldw^{(k)}$ denote the work function of $\WFA^{(k)}$\;
      Let $\Scur_k$ denote the current recommendation of $\WFA^{(k)}$\;
      \showlnlabel{line:switch}
      $\Scur_k \leftarrow \Scur_k -F^- \cup (F^+ \cap C_k)$\;
      \For{$S \subseteq C_k$} {
        $S^{\rm cons} \leftarrow S - F^- \cup (F^+ \cap C_k)$\;
        $\mathit{minDiff} \leftarrow \trans(S,S^{\rm cons}) + \trans(S^{\rm cons},S)$\;
        $\mathit{diff} \leftarrow \boldw^{(k)}[S] + \trans(S,\Scur_k) - \boldw^{(k)}[\Scur_k]$\;
        \If{ $\mathit{diff} < \mathit{minDiff}$ } 
           {Increase $\boldw^{(k)}[S]$ by $\mathit{minDiff}-\mathit{diff}$\;} 
      }
    }

  \end{myalgorithmic}
  \caption{Interface of $\algo$.}
  \label{alg:wfit}
\end{figure}

Figure~\ref{alg:wfit} shows the pseudocode for $\feedback$ based on the
previous discussion. For each part $C_k$ of the stable partition, $\feedback$
first switches the current recommendation to be consistent with the feedback
(line~\ref{line:switch}). Subsequently, it adjusts the value of
$\boldw^{(k)}[S]$ for each $S \subseteq C_k$ to enforce the 
bound (\ref{eq:sat:threshold}) on $\score(S)$.

\subsection{Maintaining Candidates Automatically}
\label{sub:dynamic_candidates_interactions}

The $\analyzeQuery$ method of $\algo$ extends the approach of
$\wfaplus$ to automatically change the stable partition as
appropriate for the current workload.
We present these extensions in the remainder of this section.
We first discuss the $\repartition$ method,
which updates $\algo$'s internal state according to a new stable partition.
Finally, we present
$\chooseCands$, which determines what that stable partition should be.

\subsubsection{Handling Changes to the Partition}
Suppose that the $\repartition$ method is given 
a stable partition $\{D_1,\dots,D_M\}$ for $\algo$ to adopt for the next queries.
We require each of the indices materialized by $\WFA$ to
appear in one of the sets $D_1,\dots,D_M$, in order to avoid 
inconsistencies between the internal state of $\algo$ and the
physical configuration.
In this discussion, we do not make assumptions about how
$\{D_1,\dots,D_M\}$ is chosen. Later in this section, we describe
how $\chooseCands$ automatically chooses the stable partition that
is given to $\repartition$.

\stitle{Unmodified Candidate Set} 
We initially consider the case where the new partition is over the same set of
candidate indices, i.e., $\bigcup_{k=1}^K C_k = \bigcup_{m=1}^M D_m$. 
The original internal state of $\algo$ corresponds to a copy of $\WFA$ for each
stable subset $C_k$. The new partition requires a new copy of $\WFA$
to be initialized for each new stable subset $D_m$. The challenge
is to initialize the work function values corresponding to $D_m$ in a meaningful
way. We develop a general initialization method that maintains an equivalence
between the work function values of $\{D_1, \dots, D_M\}$ and
$\{C_1,\dots,C_K\}$, assuming that both partitions are stable. 

We describe the reinitialization of the work function with an example. 
Assume the old stable partition is $C_1=\{a\}, C_2=\{b\}$, and 
the new stable partition has a single member $D_1=\{a,b\}$. 
Let $\boldw^{(1)}, \boldw^{(2)}$ be the work function values maintained 
by $\algo$ for the subsets $C_1,C_2$.
Let $w_n$ be the work function that considers paths in the index 
transition graph with both indices $a,b$, which represents the
information that would be maintained if $a,b$ were in the same stable subset.
In order to initialize work function values for $D_1$, we observe that
the following identity follows from the assumption that $\{C_1,C_2\}$ is a stable partition:
\[\work_n(S) 
  = \boldw^{(1)}(S \cap \{a\}) 
  + \boldw^{(2)}(S \cap \{b\})
  - \sum_{1 \leq i \leq n} \cost(q_i, \emptyset) \] 
This is a special case of Lemma~\ref{lem:partitioned_wf},
which we prove in Appendix~\ref{appendix:repartition}.
The bottom line is that it is possible to reconstruct the values of the work function
$\work_n$ using the work functions within the smaller partitions.
For the purpose of initializing the state of $\WFA$, the final sum may be
ignored: the omission of this sum increases the scores of each state $S$
by the same value, which does not affect the decisions of $\WFA$.
Based on this reasoning, our repartitioning algorithm would initialize
$D_1$ using the array $\boldx$ defined as follows:

{\small
\[ 
\begin{array}{ll}
\boldx[\emptyset] \leftarrow \boldw^{(1)}[\emptyset] + \boldw^{(2)}[\emptyset] & 
\boldx[\{a\}] \leftarrow \boldw^{(1)}[\{a\}] + \boldw^{(2)}[\emptyset] \\
\boldx[\{b\}] \leftarrow \boldw^{(1)}[\emptyset] + \boldw^{(2)}[\{b\}] & 
\boldx[\{a,b\}] \leftarrow \boldw^{(1)}[\{a\}] + \boldw^{(2)}[\{b\}] 
\end{array} \]
}%
We use an analogous strategy to initialize the work function when repartitioning
from $D_1$ to $C_1,C_2$:
 
{\small
\[\begin{array}{ll}
\boldw^{(1)}[\emptyset] \leftarrow \boldx[\emptyset] &
\boldw^{(2)}[\emptyset] \leftarrow \boldx[\emptyset] \\
\boldw^{(1)}[\{a\}] \leftarrow \boldx[\{a\}] &
\boldw^{(2)}[\{b\}] \leftarrow \boldx[\{b\}] 
\end{array} \]}%
Again, note that these assignments result in work function values that would be
different if $C_1,C_2$ were used as the stable partition for the entire workload.
The crucial point is that each work function value is distorted by the same
quantity (the omitted sum),
so the difference between the scores of any two states is preserved.

\begin{figure} 
  \begin{myalgorithmic}
    \ProcedureTitle{\FuncSty{$\repartition$}($\{D_1, \dots, D_M\}$)}%
    \KwIn{The new stable partition.}
    \tcp{Note:$\:D_1, \dots, D_M$ must cover materialized indices}
    Let $\boldw^{(k)}$ denote the work function of $\WFA^{(k)}$\;
    Let $\Scur$ denote the current recommendation of $\algo$\;
    \For{$m \leftarrow 1$ \KwTo $M$} {
      Initialize array $\boldx^{(m)}$ and configuration variable $\Tcur_m$\;
      \ForEach{$X \in 2^{D_m}$} {
      \showlnlabel{line:repartition:init}
      $\boldx^{(m)}[X] \leftarrow \sum_{k=1}^K{\boldw^{(k)}[C_k \cap X]}$\;
      \showlnlabel{line:repartition:adjustment}
      $\boldx^{(m)}[X] \leftarrow \boldx^{(m)}[X] + \trans(S_0 \cap D_m - \calC, X - \calC)$\;
      }
      $\Tcur_m \leftarrow D_m \cap \Scur$\;
    }
    Set $\{D_1, \dots, D_M\}$ as the stable partition, where $D_m$ is tracked by
    a new instance $\WFA^{(m)}$ with work function $\boldx^{(m)}$ and state $\Tcur_m$\;
  \end{myalgorithmic}
  \caption{The $\repartition$ method of $\algo$.}
  \label{alg:repartition}
\end{figure}

The pseudocode for $\repartition$ is shown in Figure~\ref{alg:repartition}. 
For each new stable subset $D_m$, the goal
is to initialize a copy of $\WFA$ with candidates $D_m$.
The copy is associated with an array $\boldx^{(m)}$ that
stores the work function values for the configurations in $2^{D_m}$. For a state
$X \subseteq D_m$, the value $\boldx^{(m)}[X]$ is initialized as the sum of
$\boldw^{(k)}[X \cap C_k]$, i.e., the work function values of the
configurations in the original partition that are maximal subsets of $X$
(line~\ref{line:repartition:init}). This initialization follows the intuition
of the example that we described previously, since the stable partition
$\{C_1,\dots,C_K\}$ implies that $X \cap C_k$ is independent from $X \cap
C_{l}$ for $k \ne l$. 
Line~\ref{line:repartition:adjustment} makes a final adjustment
for new indices in $X$, but this is irrelevant if the candidate set does not 
change (we will explain this step shortly).
Finally, the current state corresponding to $D_m$ is
initialized by taking the intersection of $\Scur$ with $D_m$. 

Overall, $\repartition$ is designed in order for the updated internal state
to select the same indices as the original state, provided that 
both partitions are stable. This property was illustrated in the example
shown earlier.
It is also an intuitive property, as two stable partitions record a 
subset of the same independencies, and hence  both allow $\algo$ 
to track accurate benefits of different configurations.
A more formal analysis of $\repartition$ would be worthwhile to
explore in future work.

\stitle{Modified Candidate Set} 
We now extend our discussion to the case where the new partition is over a
different set of candidate indices,
i.e., $\bigcup_{k=1}^K C_k \not= \bigcup_{m=1}^M D_m$. 
The $\repartition$ method (Figure~\ref{alg:repartition}) can handle this case
without modifications. The only difference is that
line~\ref{line:repartition:adjustment} becomes relevant, and it may
increase the work function value of certain configurations. 
It is instructive to consider the computation of $\boldx^{(m)}[X]$ 
when $X$ contains an index $a$ which did not previously appear in any
$C_k$ or the initial state $S_0$.
Since $a$ is a new index, it does not belong to any of the original subsets 
$C_k$, and hence the cost to materialize $a$ will not be reflected in the sum
$\sum_k \boldw^{(k)}[X \cap C_k]$. Since $\boldx^{(m)}[X]$ includes
a transition to an index set with $a$ materialized, we must add the cost to
materialize $a$ as a separate step. This idea is generalized by adding the
transition cost on line~\ref{line:repartition:adjustment}. The expression
is a bit complex, but we can explain it in an alternative form
$\trans(S_0 \cap D_m - \calC, X \cap D_m-\calC)$, which is equivalent because 
$X \subseteq D_m$. In this form, we can make an analogy to the initialization
used for the work function before the first query, for which we use 
$w_0(X) = \trans(S_0,X)$. The expression used in 
line~\ref{line:repartition:adjustment} computes the same quantity 
restricted to the indices $(D_m-\calC)$ that are new within $D_m$.


\subsubsection{Choosing a New Partition}

As the final piece of $\algo$, we present 
the method $\chooseCands$, which automatically decides
the set of candidate indices $\calC$ to be monitored by
$\WFA$, as well as the partition $\{C_1,\dots,C_K\}$ of $\calC$.
\rem{At a high level, $\chooseCands$
analyzes the workload one statement at a time, identifying interesting
indices and computing statistics on benefit interactions. These statistics
are subsequently used to compute a new stable partition,
which may reflect the addition or removal of candidate indices or changes in
the interactions among indices.}

\add{At a high level, our implementation of $\chooseCands$
analyzes the workload one statement at a time, identifying interesting
indices and computing statistics on benefit interactions. These statistics
are subsequently used to compute a new stable partition,
which may reflect the addition or removal of candidate indices or changes in
the interactions among indices. As we will see shortly, several of these steps  rely on simple, yet intuitive heuristics that we have found to work well in practice. Certainly, other implementations of $\chooseCands$ are possible, and can be plugged in with the remaining components of $\algo$.
}

The $\chooseCands$ method exposes three configuration variables that may be used to
regulate its analysis. Variable $\idxCnt$ specifies
an upper bound on the number of indices that are monitored by an instance of
$\WFA$, i.e., $\idxCnt \ge |\calC| = \sum_k |C_k|$.
Variable $\stateCnt$ specifies an upper bound on the number of
configurations tracked by $\algo$, i.e., $\stateCnt \ge \sum_k{2^{|C_k|}}$.
If the minimal stable partition does not satisfy these bounds,
$\chooseCands$ will ignore some candidate indices or some interactions
between indices, which in turn affects the
accuracy of $\algo$'s internal statistics. Variable $\histSize$ controls 
the size of the statistics recorded for past queries. Any of these variables may be set to $\infty$ in order to make the
statistics as exhaustive as possible, but this may result in high
computational overhead. Overall, these variables allow a
trade-off between the overhead of workload analysis and
the effectiveness of the selected indices.

\begin{figure}[t]
  \begin{myalgorithmic}
      \KwData{Index set $\calU \supseteq \calC$ from which to choose candidate indices;
              \newline
              Array $\idxStats$ of benefit statistics for indices in $\calU$;
              \newline
              Array $\intStats$ of interaction statistics for pairs of indices in $\calU$.}
    \FuncSep
    \sspace
    \ProcedureTitle{\FuncSty{$\chooseCands$}($q$)}%
    \KwIn{The next statement $q$ in the workload.}
    \KwOut{$D_1,\dots,D_M$, a new partitioned set of candidate indices.}
    \KwKnobs{Upper bound $\idxCnt$ on number of indices in output;
             \\ \hspace{3ex}
             Upper bound $\stateCnt$ on number of states $\sum_m 2^{|D_m|}$.
             \\ \hspace{3ex}
             Upper bound $\histSize$ on number of queries to track in statistics}

      \showlnlabel{ln:extractIndices}
      $\calU \leftarrow \calU \cup \extractIndices(q)$\;
      \showlnlabel{ln:computeIBG}
      $\mathit{IBG}_q \leftarrow \mathit{computeIBG}(q)$%
         \tcp*[l]{Based on~\cite{1687766}}
      \showlnlabel{ln:updateStats}
      $\mathit{updateStats}(\mathit{IBG}_q)$\;
         $\calM \leftarrow \{ a \in \calC ~|~ \mbox{$a$ is materialized} \}$\;
      \showlnlabel{ln:topIndices}
      $\calD \leftarrow \calM \cup \topIndices(\calU - \calM,\idxCnt - |\calM|)$\;
      \showlnlabel{ln:choosePartition}
      $\{D_1,\dots,D_M\} \leftarrow \choosePartition(\calD,\stateCnt)$\;
      \Return $\{D_1,\dots,D_M\}$\;
  \end{myalgorithmic}
  \caption{The $\chooseCands$ Method of $\algo$.}
  \label{alg:auto_stable_part}
\end{figure}

Figure~\ref{alg:auto_stable_part} shows the pseudocode of $\chooseCands$.
The algorithm maintains a large set of indices $\calU$, which
grows as more queries are seen. The goal of $\chooseCands$ is to
select a stable partition over some subset $\calD \subseteq \calU$.
To help choose the stable partition, the algorithm also maintains
statistics for $\calU$ in two arrays:
$\idxStats$ stores benefit information for individual
indices and $\intStats$ stores information about interactions between pairs of
indices within $\calU$.

Given a new statement $q$ in the
workload, the algorithm first augments $\calU$ with interesting indices 
identified by $\extractIndices$ (line~\ref{ln:extractIndices}). 
This function may be
already provided by the database system (e.g., as with IBM DB2), or it can be
implemented externally~\cite{acn:vldb00,bc:icde07}.  
Next, the algorithm computes the 
\emph{index benefit graph}~\cite{1687766} (IBG for short) of the query
(line~\ref{ln:computeIBG}). The IBG compactly encodes the costs of optimized
query plans for all relevant subsets of $\calU$.
As we discuss later, $\mathit{updateStats}$ uses the IBG to
efficiently update the benefit and interaction statistics
(line~\ref{ln:updateStats}).
The next step of the algorithm determines the new set of candidate
indices $\calD$ that should be monitored by
$\algo$ for the upcoming workload, with an invocation of
$\topIndices$ on line~\ref{ln:topIndices}. We ensure
that $\calD$ includes the currently materialized indices
(denoted $\calM$), in order to avoid overriding the materializations
chosen by $\WFA$.
Finally, $\chooseCands$ invokes $\choosePartition$
to determine the partition $D_1,\dots,D_M$ of $\calD$, and returns the result.

To complete the picture, we must describe the methodology that
$\topIndices$ and $\choosePartition$ use to
decide the new partition of indices, and the specific
bookkeeping that $\mathit{updateStats}$ does to enable this
decision.

\stitle{The $\topIndices(X,u)$ Method} 
The job of $\topIndices(X,u)$ is to choose at most
$u$ candidate indices from the set $X$ that have the highest potential benefit.

We first describe the statistics used to evaluate the potential benefit of a candidate index. For each index $a$, the $\idxStats$ array stores entries of the form $(n,\beta_n)$,
where $n$ is a position in the workload and $\beta_n$ is the \emph{maximum benefit}
of $a$ for query $q_n$.
The maximum benefit is computed as 
$\beta_n=\max_{X \subseteq \calU}{\benefit_{q_n}(\{a\},X)}$. 
The cell $\idxStats[a]$ records the $\histSize$ most recent 
entries such that $\beta_n>0$. 
These statistics are updated when $\chooseCands$ invokes $\mathit{updateStats}$ on line~\ref{ln:updateStats}. The function considers
every index $a$ that is relevant to $q$, and employs the IBG of query $q$ in
order to compute $\beta_n$ efficiently. If $\beta_n>0$ 
then $(n,\beta_n)$ is appended to $\idxStats[a]$ and the oldest entry is possibly
expired in order to keep $\histSize$ entries in total.

Based on these statistics, 
$\topIndices(X,u)$ returns a subset
$Y \subseteq X$ with size at most $u$,
which becomes the new set of indices monitored by $\algo$.
The first step of $\topIndices$ computes a
``current benefit'' for each index in $X$, which captures
the benefit of the index for recent queries.
We use $\pben_N(a)$ to denote the current benefit of $a$ after
observing $N$ workload statements, and compute this value as follows.
If $\idxStats[a]=\emptyset$ after $N$ statements,
then $\pben_N(a)$ is zero. Otherwise, let
$\idxStats[a]=(n_1,b_1),\dots,(n_L,b_L)$ such that $n_1 > \dots > n_L$. 
Then
\[ \pben_N(a) = \max_{1 \leq \ell \le L}
\frac{b_1 + \cdots + b_\ell}{N - n_\ell + 1}.
\]
For each $\ell = 1,\dots,L$, this expression computes an average benefit 
over the most recent $N - n_\ell + 1$ queries, and we take
the maximum over all $\ell$. Note that a large value of $n_\ell$ results in
a small denominator, which gives an advantage to indices with recent benefit.
This approach is inspired by the LRU-K replacement policy
for disk buffering~\cite{170081}.

The second step of $\topIndices(X,u)$ uses the 
current benefit to compute
a score for each index in $X$, and returns the
$u$ indices with the highest scores.
If $a \in X \cap \calC$ (i.e., $a$ is currently monitored by $\WFA$),
the score of $a$ is simply $\pben(a)$.
The score of other indices $b \in X - \calC$ 
is $\pben(b)$ minus the cost to materialize $b$.
This means that $b$ requires extra evidence to
evict an index in $\calC$, which helps $\calC$ be more stable.

\stitle{The $\choosePartition(\calD,\stateCnt)$ method} 
Conceptually, the stable partition models the strongest index interactions
for recent queries. We first
describe the statistics used to estimate the strength of interactions, and then
the selection of the partition.

The statistics for $\choosePartition$ are based on the 
degree of interaction $\doi_q(a,b)$ 
between indices $a,b \in \calU$ for a workload statement $q$
(Section~\ref{sec:preliminaries}). Specifically, we maintain an array $\intStats$ that is updated in the call to
$\mathit{updateStats}$ (which also updates $\idxStats$ as described earlier).
The idea is to iterate over every pair $(a,b)$ of indices in the IBG, and use
the technique of~\cite{1687766}
to compute $d \equiv \doi_{q_n}(a,b)$. 
The pair $(n,d)$ is added to $\intStats[a,b]$ if $d>0$, and only the $\mathit{histSize}$ most recent pairs are retained.

We use $\intStats[a,b]$ to
compute a ``current degree of interaction'' 
for $a,b$ after $N$ observed workload statements, denoted as $\pdoi_N(a,b)$, which is similar to the ``current benefit'' described
earlier. If $\intStats[a,b]= \emptyset$ 
then we set
$\pdoi_N(a,b)=0$. 
Otherwise, let 
$\intStats[a,b]=(n_1,d_1),\dots,(n_L,d_L)$
for $n_1 > \dots > n_L$, and
\[
\pdoi_N(a,b) = \max_{1 \leq \ell \le L}
   \frac{d_1+\cdots + d_\ell}{N-n_\ell+1}.
\]

To compute the stable partition, we conceptually
build a graph where vertices correspond to indices and edges
correspond to pairs of interacting indices. Then a stable partition is
a clustering of the nodes so that no edges exist between clusters. In the
context of $\chooseCands$, we are interested in partitions 
$\{P_1, \dots, P_M\}$ such that
$\sum_m 2^{|P_m|} \le \stateCnt$. Since there may exist no stable partition
that obeys this bound, our approach is to ignore interactions until a feasible
partition is possible. This corresponds to dropping edges from the conceptual
graph, until the connected components yield a suitable clustering of the nodes.

An important question is which interactions to ignore.
Our strategy is to minimize the error that the partition introduces in
the formula for query cost (\ref{eq:querycost}), which is the basis of
all statistics tracked by $\algo$. It is straightforward to show that
the error in (\ref{eq:querycost}) is bounded by the sum of $\doi$ values
for ignored interactions. Hence we define the loss of a partition
$P = \{P_1,\dots,P_M\}$ as
\[
\loss(P) = \sum_{i < j} \sum_{a \in P_i} \sum_{b \in P_j} \pdoi_N(a,b).
\]
In the graph-based interpretation,
this corresponds to the sum of edge weights for edges that cross clusters.

\begin{figure}[tb]
  \newcommand\bestSoln{\mathit{bestSoln}}
  \newcommand\bestLoss{\mathit{bestLoss}}
  
  \begin{myalgorithmic}
    \FunctionTitle{\FuncSty{$\choosePartition$}($\calD$, $\stateCnt$)}
    \KwIn{Indices $\calD$ to partition;
      \\ \hspace{3ex}
      Bound $\stateCnt \ge \sum 2^{|P_m|}$ for the output $\{P_1, \dots, P_M\}$}
      $\bestSoln \leftarrow \emptyset$;~~$\bestLoss \leftarrow \infty$\;
      \tcp{Try a baseline partition that is similar to the current one} 
      Let $\{C_1,\dots,C_K\}$ denote the current partition and $\calC = \bigcup_k C_k$\;
      Initialize $\{C'_1,\dots,C'_K\}$ by removing $\calC - \calD$ from $\{C_1,\dots,C_K\}$\;
      $P \leftarrow \{C'_1,\dots,C'_K\}$ \;
      \lForEach{$a \in \calD - \calC$}{Add $\{a\}$ to $P$\;}
      \If{\emph{$P$ is feasible (i.e., satisfies the bound $\stateCnt$)}}{
        $\bestSoln \leftarrow P$;~~$\bestLoss = \loss(P)$ \;
      }
      \tcp{Try additional random partitions}
      \For{$i \leftarrow 1$ \KwTo $\randCnt$}{
        $P \leftarrow $ a partition of $\calD$ in singletons\;
        \While{true} {
          Let $\{P_1,\dots,P_M\}$ denote the contents of $P$\;
          $E\!\leftarrow\! \{ \! \{ \! P_i,P_j \! \} ~|~ 
               \loss(\!\{\!P_i,P_j\!\}\!) \!>\!0 
               \land \mbox{feasible to merge $P_i,P_j$}\!\}$\;
          \lIf{$E = \emptyset$}{ break\; }
          \ElseIf{$E_1 \equiv \{ \{P_i,P_j\} \in E ~~|~~ 1=|P_i|=|P_j|\} \ne \emptyset$} {
            Choose random $\{P_i,P_j\} \in E_1$ 
            with probability 
            proportional to $\loss(\{P_i,P_j\})$ \;
          }
          \Else {
            Choose $\{P_i,P_j\} \in E$ 
            with probability proportional to 
            $\loss(\{P_i,P_j\})/(2^{|P_i|+|P_j|}-2^{|P_i|}-2^{|P_j|})$ \;
          }
          $P \leftarrow \mbox{result of merging $P_i,P_j$ in $P$}$ \;
        }
\eat{ 
        \While{true} {
          Let $\{D_1,\dots,D_M\}$ be the contents of $P$\;
          $E \leftarrow \{ (D_i,D_j) ~|~ \loss(\{D_i,D_j\}) > 0 \land |D_i|=|D_j|=1\}$\;
          \lIf{$E = \emptyset$}{break}\\
          \lElse{
            Choose random $(D_i,D_j)$ from $E$ to merge with probability proportional to $\loss(\{D_i,D_j\})$ \;
          }
        }
        \While{true}{
          Let $\{D_1,\dots,D_M\}$ be the contents of $P$\;
          $E \! \leftarrow \! \{ (D_i,D_j) \,|\, \loss(\{D_i,D_j\}) \!>\!0 \land \mbox{feasible to merge $D_i,D_j$}\}\!$ \;
          \lIf{$E = \emptyset$}{break}\\
          \lElse {
            Choose random $(D_i,D_j)$ from $E$ to merge with probability proportional to $\loss(\{D_i,D_j\})/(2^{|D_i|+|D_j|}-2^{|D_i|}-2^{|D_j|})$ \;
          }
        }
} 
        \If{$\loss(P) < \bestLoss$} {
          $\bestSoln \leftarrow P$; $\bestLoss \leftarrow \loss(P)$ \;
        }
      }
      \Return $\bestSoln$\;
  \end{myalgorithmic}
  \caption{Function $\choosePartition$.}
  \label{alg:choosePartition}
\end{figure}

Figure~\ref{alg:choosePartition} shows the pseudocode for function $\choosePartition$ that computes the new stable partition. The goal is to return a feasible partition that minimizes loss. 
We employ a randomized approach that finds several feasible partitions and
returns the one with the least loss. As a baseline solution, the function
considers the existing stable partition, augmented with singleton parts for the
new indices in $\calD$. It then performs $\randCnt$ randomized
iterations, where $\randCnt$ is a parameter of the algorithm.
Each iteration has two stages. The first stage simply merges singleton
sets that exhibit a high degree of interaction. The pair $(\{a\},\{b\})$
to merge is chosen randomly with weight proportional to
$\pdoi_N(a,b)$.
The second stage is similar, with a different weighting
scheme. Given two sets $A$ and $B$, it assigns a weight proportional to
$\sum_{a \in A} \sum_{b \in B}{\pdoi_N(a,b) / (2^{|A|+|B|}-2^{|A|}-2^{|B|})}$. 
The normalization accounts for the additional
number of configurations that will result from merging $A$ and $B$.
Essentially, the weight assigned to $(A,B)$ represents the increase in 
$\loss(P)$ per
additional state tracked by $\algo$, and hence the merging favors small
sets whose indices have strong interactions. The function returns
the best partition found across all iterations. 

This concludes the final piece of the $\algo$ algorithm.
As a final note, observe that the methods of $\algo$ use strategies that
are quite orthogonal. In particular, the method
$\repartition$ does not depend on the specific heuristics
that $\chooseCands$ uses to determine the candidate indices.
With this design, it is straightforward to substitute
$\chooseCands$ with alternate strategies for candidate selection
and partitioning. There is a broad design space for this component
of $\algo$, and this would be an interesting direction for future work.

  \section{Experimental Study} 
\label{sec:experimental_study}

In this section, we present an empirical evaluation of $\algo$ using a
prototype implementation that works as middleware on top of an existing DBMS.
The prototype, written in Java, intercepts the SQL queries and analyzes them to
generate index recommendations. The prototype requires two services from the
DBMS: access to the what-if optimizer, and an implementation of the \linebreak
$\extractIndices(q)$ method (line~\ref{ln:extractIndices} in
Figure~\ref{alg:auto_stable_part}). 
\add{This design makes the prototype easily portable, as these services are
common primitives found in index advisors~\cite{vzzls:icde00,acn:vldb00}.}

We conducted experiments using a port of the prototype to the IBM DB2 Express-C DBMS. The port uses DB2's design advisor~\cite{vzzls:icde00} to provide what-if optimization and $\extractIndices(q)$. Unless otherwise noted, we set the parameters of $\algo$ as follows: $\idxCnt=40$, $\stateCnt=500$, and $\histSize=100$. All experiments were run on a machine with two dual-core 2GHz Opteron processors and 8GB of RAM.

\eat{
We also completed a port of our prototype over the PostgreSQL DBMS. In this case, we
extended the PostgreSQL query optimizer with a what-if interface and added our own
implementation of \linebreak $\extractIndices(q)$ following the heuristics presented
in~\cite{bc:icde07}. We conducted only a limited number of experiments using this
port, in order to gauge the generality of our findings. We report on these
experiments when possible.
}

\subsection{Methodology}

\stitle{Competitor Techniques} We compare $\algo$ empirically against two competitor algorithms. The first
algorithm, termed $\BC$, is an adaptation\footnote{The original algorithm was
developed in the context of MS SQL Server. Some of its components do not have
counterparts in DB2.} of the state-of-the-art online tuning algorithm of
Bruno and Chaudhuri~\cite{bc:icde07}. $\BC$ analyzes the workload using ideas
similar to $\algo$, except that it always employs a stable partition
corresponding to full index independence, i.e., each part contains a single
index. After a query is analyzed, $\BC$ heuristically adjusts the measured
index benefits to account for specific types of index interactions. The
principled handling of index interactions is a major difference between $\algo$
and $\BC$.

The second alternative is $\OPT$, which has full knowledge of the workload and generates the optimal recommendations that minimize total work. $\OPT$ provides a baseline for the best-case performance of any online index recommendation algorithm. 

In order to make a meaningful comparison between these algorithms, some of our experiments use
a fixed set of candidates $\calC$ and stable partition $\{C_1,\dots,C_K\}$
throughout the workload.
In this way, the algorithms select their recommendations from the same
configuration space, and our experiments focus on the recommendation
logic. This approach requires a simplification of $\algo$ so that
$\chooseCands$ always returns $\{C_1,\dots,C_K\}$. Our final experiment
compares the simplified version of $\algo$ to the full version that
allows $\chooseCands$ to modify the stable partition throughout the workload.

\stitle{Data Sets and Workloads} We base the experimental study on an index
tuning benchmark that we introduced in our previous work~\cite{Schnaitter:2009fk}. \add{The benchmark is designed to stress test the effectiveness of online tuning algorithms, and it has already been used to compare existing methods.} The benchmark simulates a system hosting
multiple databases using the  synthetic data sets TPC-C, TPC-H and TPC-E and
the real-life data set NREF, with a total of 2.9GB of base-table data. We note that the database size is not a crucial statistic for our study, as we evaluate the performance of index-tuning algorithms using the optimizer's cost model (see discussion below).

\add{We use the \emph{complex} workload defined by the benchmark, which includes
SQL query and update statements. Each statement involves a varying number of joins
and selection predicates of mixed selectivity.
The following is an example query from the workload:}

{\scriptsize\tt
\begin{tabbing}
SELECT count(*)\\
FROM tpce.security table1, tpce.company table2,\\
\ \ \ \ \ tpce.daily\_market table0\\
WHERE table1.s\_pe BETWEEN 63.278 AND 86.091\\
\ \ AND table1.s\_exch\_date BETW\=EEN '1995-05-12-01.46.40'\\
                                 \>AND '2006-07-10-01.46.40'\\
\ \ AND table2.co\_open\_date BETWEEN '1812-08-05-03.21.02'\\
                                \>AND '1812-12-12-03.21.02'\\
\ \ AND table1.s\_symb = table0.dm\_s\_symb\\
\ \ AND table2.co\_id=table1.s\_co\_id
\end{tabbing}}

\noindent
\add{And the following is an example update:}

{\scriptsize\tt
\begin{tabbing}
UPDATE tpch.lineitem\\
SET l\_tax = l\_tax + RANDOM\_SIGN()*0.000001\\
WHERE l\_extendedprice BETWEEN 65522.378 AND 66256.943
\end{tabbing}}

\noindent
\add{This update statement uses a user-defined function
{\small\tt RANDOM\_SIGN()} which randomly returns $1$ or $-1$ with
equal probability.}
\rem{The workload comprises randomly generated SQL query and update statements.
We use the \emph{complex} statement template 
of the original study that generates queries with several joins and selection
predicates of mixed selectivity.}

The workload is separated in
eight consecutive phases. Each phase comprises 200 statements and favors statements on specific data sets, thus requiring a different set of indices for effective tuning. Adjacent phases
overlap in the focused data sets and also differ in the relative frequency of
updates and queries. (See~\cite{Schnaitter:2009fk} for further details on
data and SQL statements.) The specific workload is a difficult use case for index tuning due to the mix of updates and queries and the alternation of phases. In fact, the DB2 index advisor was unable to recommend a beneficial index configuration for the whole workload, even with an infinite storage budget for indices.
(We obtained similar experimental results with workloads of lower query complexity.)

\stitle{Performance Metrics} We measure the performance of an online algorithm $A$ using $\totwork(A,Q_n,V)$ for the previously described workload and some feedback stream $V$. The definition of $V$ depends on the experiment and is detailed when we present the results. As in previous studies on index tuning~\cite{bc:icde07,DBLP:journals/pvldb/BrunoC08,Schnaitter:2009fk}, the total work metric is evaluated using the optimizer's cost model. The goal is to isolate the performance of $A$ from any cost-estimation errors, e.g., due to insufficient data statistics or faulty cost models.

In all experiments, the we measure the performance of $A$ as
$\totwork(\OPT,Q_n,V)/\totwork(A,Q_n,V)$,
which indicates the performance of $A$ relative to the optimal recommendations
of $\OPT$.
We note that the $\OPT$ can have very different recommendation
schedules for $Q_n$ and $Q_{n+1}$ respectively, whereas $A$'s recommendation
schedule for $Q_{n+1}$ is an extension of the schedule for $Q_n$.

We also report the overhead of algorithm $A$ in terms of two components: the number of what-if optimization calls, and the remainder of the overhead as absolute wall-clock time.
The reason for this separation is that the efficiency of the what-if optimizer is somewhat independent of the tuning algorithm. Indeed, techniques for very fast what-if optimization~\cite{DBLP:conf/sigmod/BrunoN08} can reduce substantially the overhead of any tuning task.

\stitle{Generating the Fixed Stable Partition} 
As explained above, we choose a fixed stable partition $\{C_1,\dots,C_K\}$
to be used by the competing algorithms. We developed an automated method to compute this partition in a way that captures the most relevant indices and interactions in the entire workload. Specifically, we first obtain a large set of interesting indices $\calU$
by invoking DB2's index advisor on
\add{the read-only portion of the workload} with an infinite space budget
\add{(as mentioned earlier, the index advisor would not recommend
any indices to create for the entire workload)}.
We then choose a subset $\calC \subseteq \calU$ and a partition
of $\calC$, using an offline variation of the $\chooseCands$ algorithm.
The only change to $\chooseCands$ is to compute an average of the
benefit and degree of interaction over the entire workload
(rather than a suffix), and use these measurements as the criteria
for the top indices and stable partition.
\add{For the workloads in our experiments, $\calU$ contained roughly 300 indices,
and the size of the stable partition depended on the parameter settings of $\algo$.}

\begin{figure*}[ht]
  \centering
  \begin{tabular}{ccc}
    \includegraphics[scale=0.16]{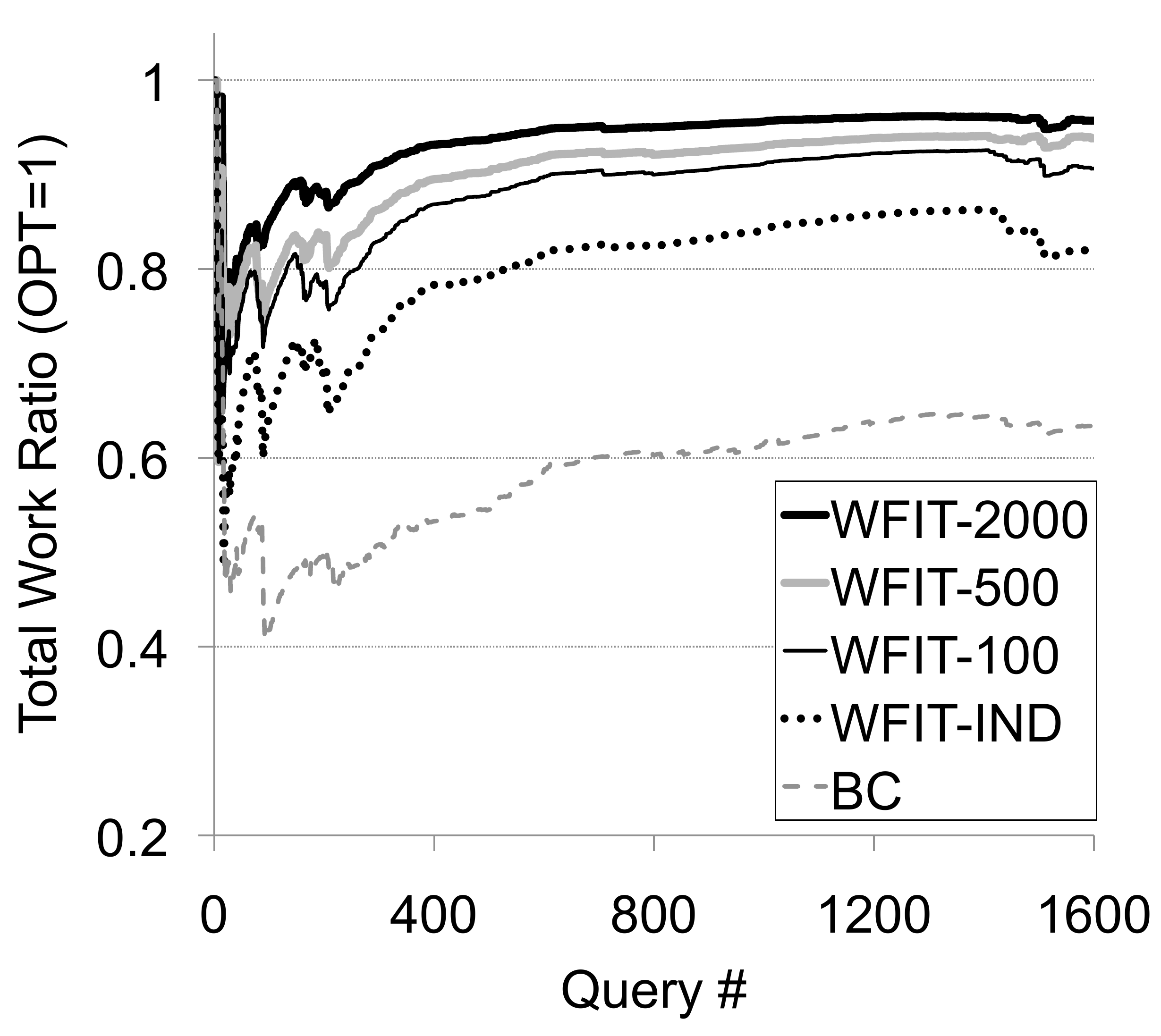} & 
    \fspace 
    \includegraphics[scale=0.16]{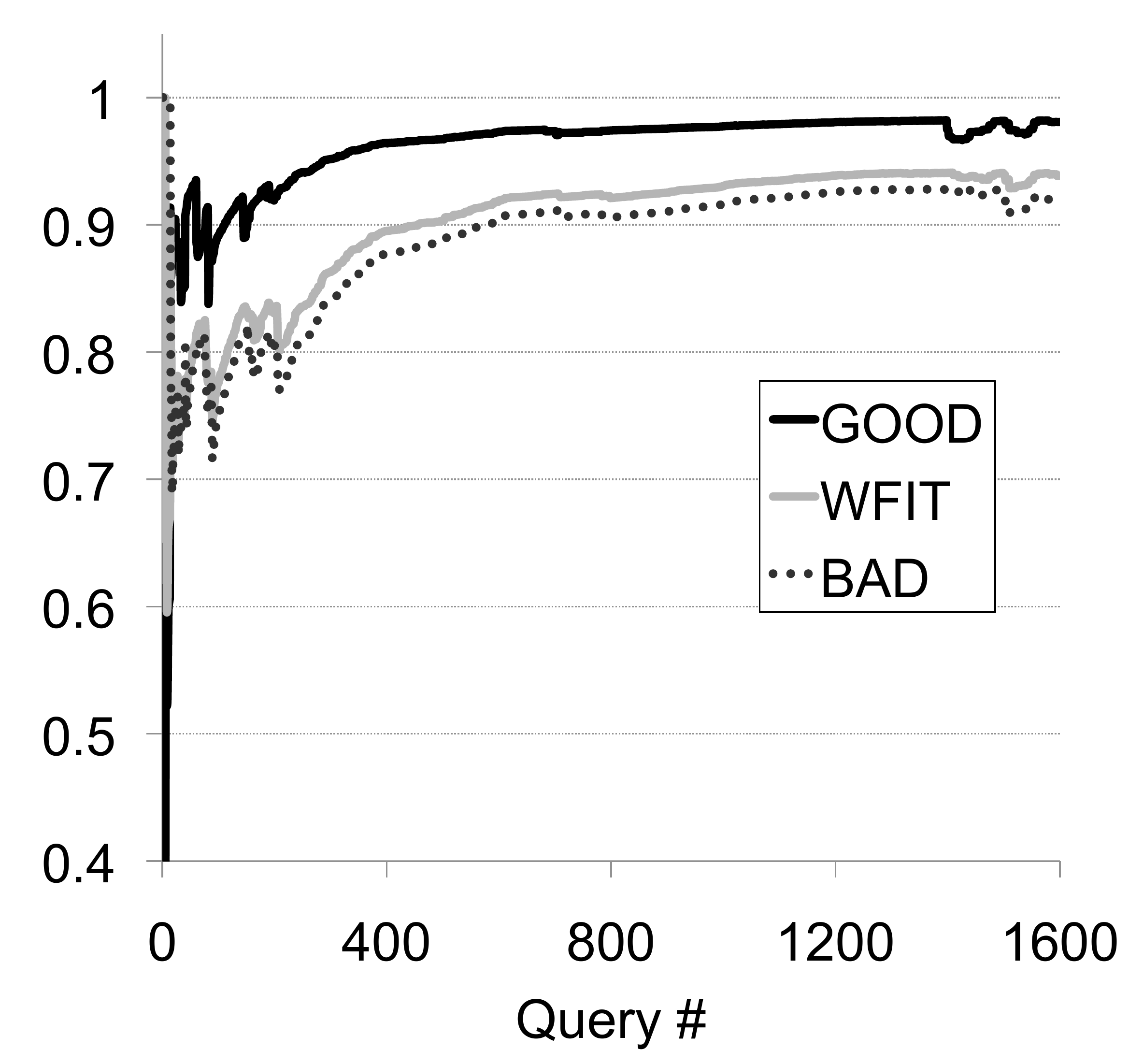} & 
    \fspace 
    \includegraphics[scale=0.16]{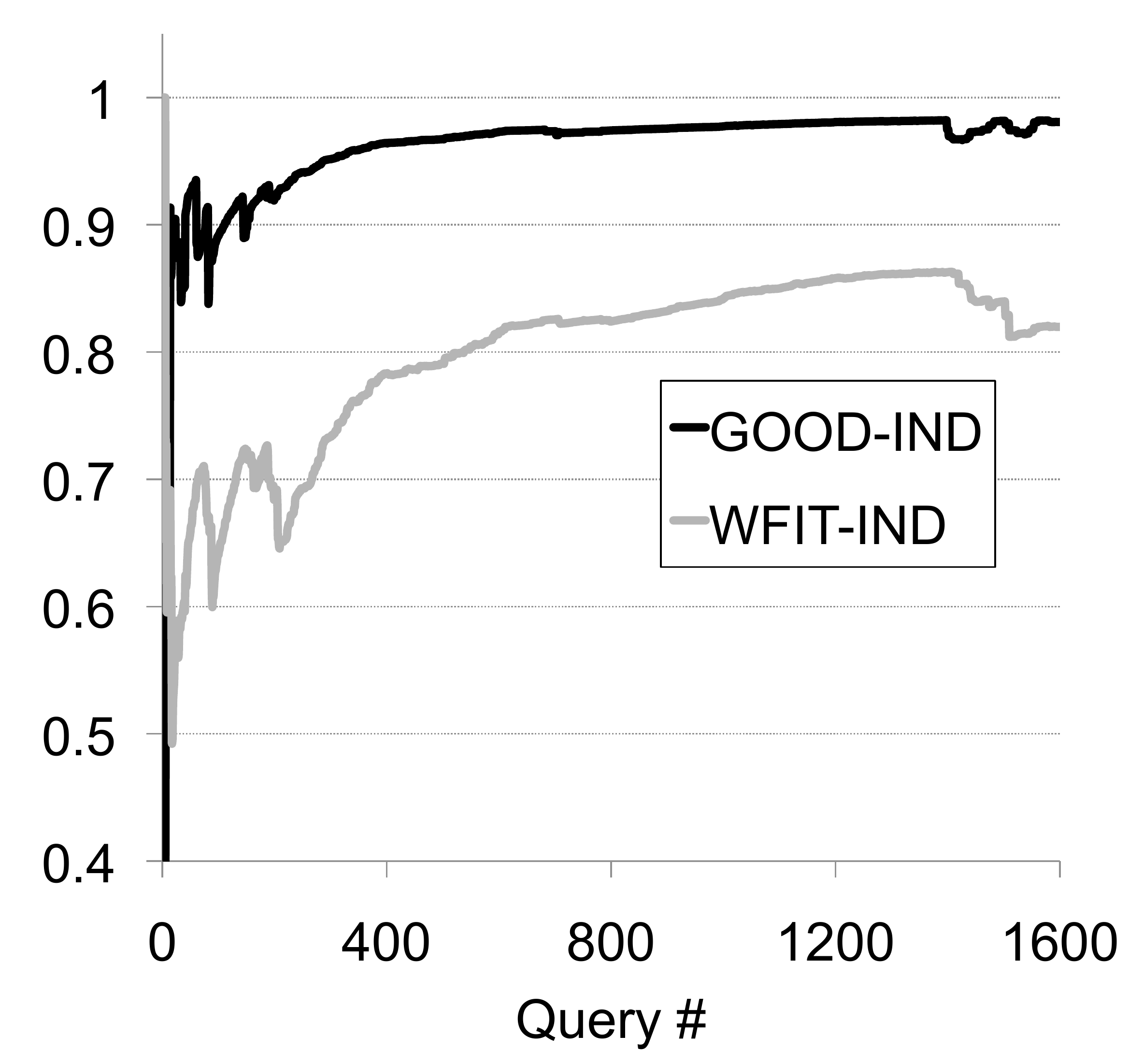}\\
    \tspace
    \captioncell{0.25\hsize}{Baseline performance \newline evaluation.}{fig:experiment1} &
    \captioncell{0.31\hsize}{Effect of DBA's feedback.}{fig:experiment23} &
    \captioncell{0.27\hsize}{Effect of DBA's feedback under independence assumption}{fig:experiment24}
  \end{tabular}
\end{figure*}

\begin{figure*}[ht]
  \centering
  \vpar
  \begin{tabular}{cc}
    \sspace
    \includegraphics[scale=0.16]{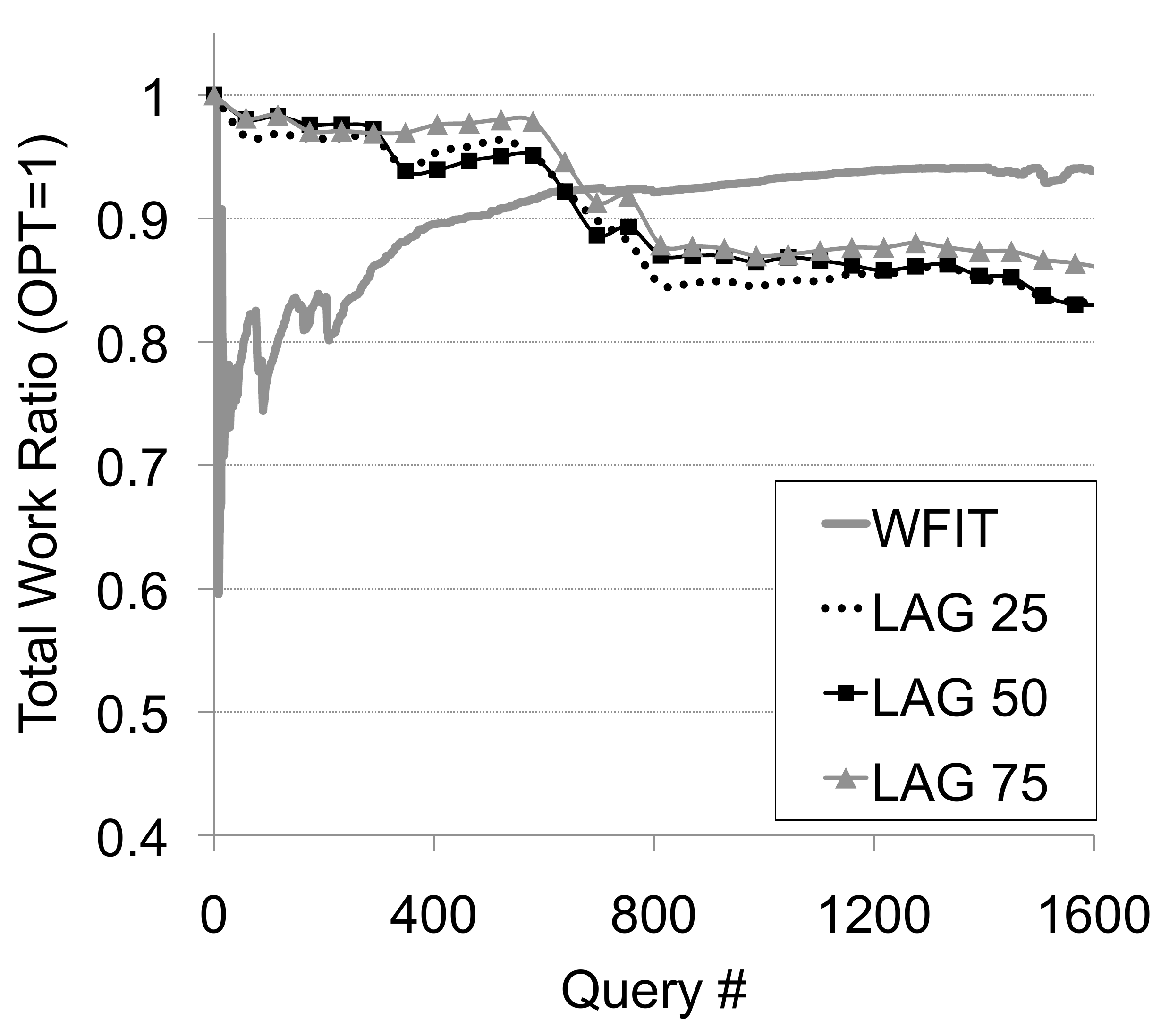} &
    \sspace 
    \includegraphics[scale=0.16]{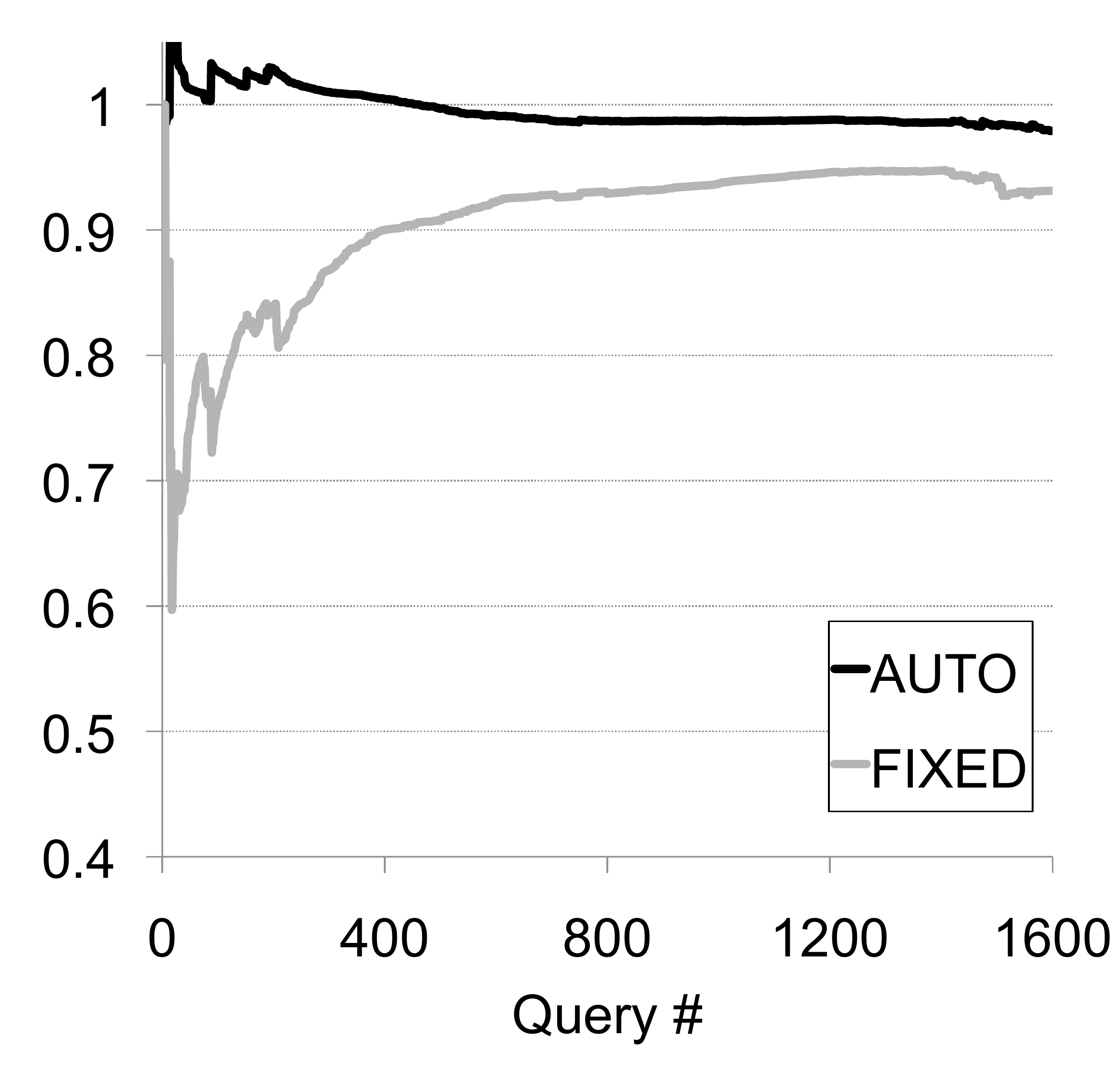}\\
    \captioncell{0.33\hsize}{Effect of delayed responses.}{fig:experiment3} &
    \captioncell{0.27\hsize}{Automatic maintenance \newline of stable partition.}{fig:experiment4}
  \end{tabular}
  \sspace
\end{figure*}

\subsection{Results}

\stitle{Baseline Performance} We begin with a baseline experiment where the stable partition is fixed and no feedback is provided ($V=\emptyset$). In this setting, $\algo$ becomes equivalent to $\wfaplus$ (Section~\ref{sec:wfa}) and the measured performance reflects the effectiveness of the index recommendation logic. It also becomes possible to make a meaningful comparison to $\BC$, which does not support feedback.

Figure~\ref{fig:experiment1} shows the normalized performance metrics for
$\algo$ and $\BC$. For $\algo$ we chart three curves that correspond to three
different settings of 2000, 500, and 100 for the $\stateCnt$ parameter
of the stable partition.
A high value corresponds to a more detailed stable partition that provides more
information to $\algo$ but also increases its overhead. (The complexity of
$\algo$ grows quadratically with $\stateCnt$.)
\add{Figure~\ref{fig:experiment1} also includes a fourth curve labeled
WFIT-IND, which corresponds to a variant of $\algo$ that considers
all indices to be independent. In other words, this version of the
algorithm assumes $\doi_q(a,b) = 0$ for all indices and queries, which
means that each index is in a separate singleton part. This version
of $\algo$ would not be used in practice, but we show its performance
in order to see the value of analyzing index interactions.}

As shown, the quality of recommendations degrades gracefully as
$\stateCnt$ decreases from 2000 down to 100, with
the overall difference remaining small throughout.
\add{The drop in performance is more significant for WFIT-IND, where all index
interactions are ignored.}
\add{We performed experiments with higher settings of
$\stateCnt$, up to $10000$, but we omit the results, as there
was very little difference compared to $\stateCnt = 2000$.}
Essentially, the results show that $\algo$ can generate effective
recommendations as long as the stable partition captures the important
interactions among the candidate indices.

Another observation from Figure~\ref{fig:experiment1} is that $\algo$'s performance comes very close to the algorithm that has complete knowledge of the workload. The difference is less than 10\% at the end, which is very significant if one considers the complex mix of updates and join queries in the workload. It is interesting to examine this empirical performance against the theoretical competitive ratio stated in Section~\ref{sec:wfa}. For this particular experiment, there are 8 indices in the biggest part of the stable partition and hence the performance of $\algo$ should always be within a factor of $2^{8+1}-1$ of optimal. As shown by the results, $\algo$'s performance can be much better compared to this worst-case bound. 

Finally, Figure~\ref{fig:experiment1} shows that $\algo$ outperforms $\BC$ by a significant margin. The difference becomes substantial after the initial statements in the workload, and by the end $\algo$ (without the independence assumption) attains $>$90\% of the performance of $\OPT$ compared to 65\% for $\BC$. The difference shows that $\algo$'s principled handling of index interactions is more effective than the heuristics used by $\BC$.
\add{In fact, the results show that even WFIT-IND outperforms $\BC$ on this
workload. This could be due in part to the fact that our adaptation of $\BC$
is implemented outside the DBMS, and the original design of $\BC$ may be better
suited for an internal implementation that is closely integrated with the
query optimizer.}
\rem{Experiments with different workloads and with our PostgreSQL prototype yielded similar trends in terms of the quality of $\algo$'s recommendations and its relative performance compared to $\BC$.}

\stitle{Overhead} For the same experiment, the Java implementation of $\algo$ on top of DB2 required 300ms on average to analyze each query and generate the recommendations. This magnitude of overhead is acceptable if one considers the much higher query execution cost and the savings obtained from having the right indices materialized. Still, overhead can be reduced substantially with a careful implementation inside the DBMS, or by switching to a lower value for $\stateCnt$. For instance, setting $\stateCnt=100$ will not affect significantly the quality of recommendations (see Figure~\ref{fig:experiment1}) but it can reduce the overhead by a factor of 25. A different solution is to do the analysis in a separate machine (e.g., the DBA's workstation) without any impact on normal query evaluation. 

Regarding the number of what-if optimizations, $\algo$ averaged between 5 and 100 calls per query close to the start and end of the experiment respectively. The number of what-if calls is directly correlated with the number of candidate indices that are mined from the workload. A different implementation of $\algo$ could constrain the latter, but the experimental results of Bruno and Nehme~\cite{DBLP:conf/sigmod/BrunoN08} suggest that it is possible to perform 100 what-if calls per query while keeping up with the flow of the workload. 

\stitle{The Effect of Feedback} The next set of experiments evaluates $\algo$'s feedback mechanism (Section~\ref{sub:feedback}), one of the core features of the semi-automatic tuning paradigm. 

We examine the performance of $\algo$ for two contrasting models of DBA feedback. The first model, represented with a feedback input $V_{\GOOD}$, represents ``good'' feedback where
the DBA casts a positive 
(resp. negative) vote for index $a$ at point $n$ in the workload if $\OPT$
creates (resp. drops) $a$ after analyzing query $n$. The idea is to model a
prescient DBA who can use votes to guide $\algo$ toward the optimal design.
We also create a ``bad'' feedback input, denoted as $V_{\BAD}$, as the mirror
image of good feedback, i.e., we replace the positive votes with negative votes
and vice versa.

Figure~\ref{fig:experiment23} shows the performance of $\algo$ for $V=V_{\GOOD}$ and $V=V_{\BAD}$. As a baseline, we include a run of $\algo$ without feedback, i.e., $V=\emptyset$.
The results show that the feedback mechanism works intuitively. The useful
feedback improves the performance of the baseline and pushes it closer to the
optimal algorithm. $\algo$ does not exactly match the performance of $\OPT$, since the
latter computes its recommendations using much more detailed information
(recall that $\algo$ uses a fixed stable partition with $\stateCnt=500$). The
bad feedback causes a degradation of performance, as expected, but $\algo$ is
still able to output effective recommendations and remain above 90\% of optimal by the end of the workload. The key point is that
$\algo$ initially biases its recommendations according to the erroneous
feedback, but it is able to recover based on the subsequent analysis of the
workload.

\add{It is also interesting to examine the effect of feedback in the modified
version of $\algo$ which assumes all indices are independent.
This experiment models an interesting scenario for the usefulness of
semi-automatic tuning, as the assumption of index independence can introduce
significant errors in $\algo$'s internal statistics on index benefits, and hence
DBA feedback can have a significant effect on the quality of the generated
recommendations.
Figure~\ref{fig:experiment24} shows the result of providing feedback
as $V_{\GOOD}$ for the WFIT-IND algorithm. (We omit results that combine
WFIT-IND with the ``adversarial'' feedback $V_{\BAD}$, since such a scenario
would stray too far from what would be seen in practice, and the results would
have little meaning.)
The results show that the DBA's feedback can still improve the quality of
the recommendations significantly, despite the fact that $\algo$ has very
inaccurate internal statistics.
}

\stitle{Delayed Feedback} The previous experiments assumed that the DBA accepts the recommendation of $\algo$ after each query. In contrast, the next experiments evaluate the effect of delayed feedback, which is what we expect to see in practice. We model this scenario with a feedback input $V_T$, where the DBA requests and accepts the current recommendation of $\algo$ every $T$ queries. This feedback renews the ``lease'' of the current recommendation, which in turn delays $\algo$ from switching to a potentially better recommendation. Hence, some degradation in performance is possible.

The results of this experiment are shown in Figure~\ref{fig:experiment3}.
The first curve shows the performance for $T = 1$, which grants full autonomy to $\algo$. The other curves show the result of increasing the delay $T$ to $25$, $50$, and $75$. There is clearly a loss in overall performance when the responses of the DBA are delayed. 
At the end of the workload, the performance with $T > 1$ is around
85\% of optimal, which is below the 95\% level achieved by $\algo$
without the lag. A close examination of the results reveals that most 
indices are beneficial only for short windows of the workload,
due to intervening updates that make indices expensive to maintain. This
aspect of the workload makes the delayed responses particularly detrimental,
and reflects our choice of this workload as a ``stress test'' for $\algo$.
However, it is important to observe that the performance does not continue to
degrade as the length of the lag increases.
We limited the lag to 75 queries in order to avoid a lag that spanned a large
portion of the phase length of 200 queries. 
In general, the results suggest that
semi-automatic interface can provide robust recommendations even when the lag
is significant compared to the phase length.

\stitle{Automatic Maintenance of Stable Partition}
The final set of experiments examines the performance of $\algo$ when
$\chooseCands$ is used to maintain the stable partition automatically, as described in
Section~\ref{sub:dynamic_candidates_interactions}. In this case, the stable
partition may change over time, which causes 
$\repartition$ 
to be
invoked. We compare this approach to the variation of $\algo$ with a fixed
stable partition.

Figure~\ref{fig:experiment4} shows the performance of $\algo$ with a fixed
stable partition and with automatic maintenance of candidates, labeled
FIXED and AUTO, respectively. 
We see an overall improvement in the performance using $\chooseCands$
to maintain the indices and interactions on-the-fly. Overall, $\chooseCands$ mined about 300 candidate indices from the workload, and changed the stable partition 147 times over the course of the experiment
(although many of the calls to $\repartition$ only made minor
changes to the modeled interactions). The observed performance clearly
validates the ability of $\repartition$ to update the internal state
of $\algo$ in a meaningful way. 
We also observe that the performance slightly exceeds $\OPT$
in the earlier queries, which are mostly read-only statements.
This is due to the fact that the automatic maintenance of the stable partition
allows $\algo$ to specialize the choice of indices for each phase, whereas
$\OPT$ is limited to one set of candidates for the workload.


  \section{Conclusions}

We introduced the novel paradigm of semi-automatic index tuning, and its realization in the $\algo$ algorithm. $\algo$ leverages and extends principled methods from online computation. Experimental results validate its numerous advantages over existing techniques, and the feasibility of semi-automatic tuning in practice.

  {
  \bibliographystyle{abbrv}
  \bibliography{main}

\begin{thebibliography}{10}

\bibitem{acn:vldb00}
S.~Agrawal, S.~Chaudhuri, and V.Narasayya.
\newblock {Automated Selection of Materialized Views and Indexes for SQL
  Databases}.
\newblock In {\em VLDB}, 2000.

\bibitem{1142549}
S.~Agrawal, E.~Chu, and V.~Narasayya.
\newblock Automatic physical design tuning: workload as a sequence.
\newblock In {\em SIGMOD}, 2006.

\bibitem{Borodin:1998ec}
A.~Borodin and R.~El-Yaniv.
\newblock {\em Online Computation and Competitive Analysis}.
\newblock Cambridge University Press, 1998.

\bibitem{bls:jacm92}
A.~Borodin, N.~Linial, and M.~E. Saks.
\newblock An optimal on-line algorithm for metrical task system.
\newblock {\em J. ACM}, 39(4), 1992.

\bibitem{bc:icde07}
N.~Bruno and S.~Chaudhuri.
\newblock An online approach to physical design tuning.
\newblock In {\em ICDE}, 2007.

\bibitem{DBLP:journals/pvldb/BrunoC08}
N.~Bruno and S.~Chaudhuri.
\newblock Constrained physical design tuning.
\newblock {\em PVLDB}, 1(1), 2008.

\bibitem{5447800}
N.~Bruno and S.~Chaudhuri.
\newblock Interactive physical design tuning.
\newblock In {\em ICDE}, 2010.

\bibitem{DBLP:conf/sigmod/BrunoN08}
N.~Bruno and R.~V. Nehme.
\newblock Configuration-parametric query optimization for physical design
  tuning.
\newblock In {\em SIGMOD}, 2008.

\bibitem{Hu:2008:QVQ:1454159.1454209}
L.~Hu, K.~A. Ross, Y.-C. Chang, C.~A. Lang, and D.~Zhang.
\newblock Queryscope: visualizing queries for repeatable database tuning.
\newblock {\em Proc. VLDB Endow.}, 1:1488--1491, August 2008.

\bibitem{mwd:ssdbm09}
T.~Malik, X.~Wang, D.~Dash, A.~Chaudhary, A.~Ailamaki, and R.~C. Burns.
\newblock Adaptive physical design for curated archives.
\newblock In {\em SSDBM}, 2009.

\bibitem{170081}
E.~J. O'Neil, P.~E. O'Neil, and G.~Weikum.
\newblock The {LRU}-{K} page replacement algorithm for database disk buffering.
\newblock In {\em SIGMOD}, 1993.

\bibitem{1325974}
S.~Papadomanolakis, D.~Dash, and A.~Ailamaki.
\newblock Efficient use of the query optimizer for automated physical design.
\newblock In {\em VLDB}, 2007.

\bibitem{slss:smdb07}
K.-U. Sattler, M.~L{\"{u}}hring, I.~Geist, and E.~Schallehn.
\newblock Autonomous management of soft indexes.
\newblock In {\em SMDB}, 2007.

\bibitem{Polyzotis:2007lr}
K.~Schnaitter, S.~Abiteboul, T.~Milo, and N.~Polyzotis.
\newblock On-line index selection for shifting workloads.
\newblock In {\em SMDB}, 2007.

\bibitem{Schnaitter:2009fk}
K.~Schnaitter and N.~Polyzotis.
\newblock {A Benchmark for Online Index Selection}.
\newblock In {\em SMDB}, 2009.

\bibitem{1687766}
K.~Schnaitter, N.~Polyzotis, and L.~Getoor.
\newblock Index interactions in physical design tuning: modeling, analysis, and
  applications.
\newblock {\em Proc. VLDB Endow.}, 2(1), 2009.

\bibitem{vzzls:icde00}
G.~Valentin, M.~Zuliani, D.~C. Zilio, G.~Lohman, and A.~Skelley.
\newblock {DB2} advisor: An optimizer smart enough to recommend its own
  indexes.
\newblock In {\em ICDE}, 2000.

\end{thebibliography}
  }
  
  \appendix
  
  \section{Competitive Analysis}
\label{appendix:competitive}

This section provides proofs of our results on the competitive ratio of
$\WFA$ (Theorem~\ref{thm:wfa_competitive}) and
$\wfaplus$ (Theorem~\ref{thm:wfit_competitive}).
The proof starts with two technical lemmas that lead to a central
result given in Theorem~\ref{thm:wfa_cost}. This theorem leads immediately
to Theorem~\ref{thm:wfa_competitive}, and it is also used to prove
Theorem~\ref{thm:wfit_competitive} with a bit more machinery.

We begin with notation.
Fix the workload $Q$
of $N$ statements and denote the $i$-th query as $q_i$.
In the context of this analysis, the algorithms choose
recommendations that are subsets of a fixed set of 
candidate indices $\calC$. Prior to observing any queries,
the materialized set of indices is some set $S_0 \subseteq \calC$.

We introduce a formal expression for the transition cost $\trans(X,Y)$:
\[ \trans(X,Y) = \sum_{a \in Y-X} \trans^+(a) + \sum_{a \in X-Y} \trans^-(a) \]
where $\trans^+(a)$ and $\trans^-(a)$ denote 
respectively the cost of creating and dropping index $a$.
We use the shorthand notation
\[ \WFA_n = \totwork(\WFA,Q_n,\emptyset) \]
for the total work of $\WFA$ on a prefix $Q_n$ of the workload. We 
define the shorthand $\wfaplus_n$ and $\OPT_n$ similarly.

For each query $q_i$ we fix a set $S^*_i$ that minimizes the
cost of $q_i$:
\[ S^*_i = \arg\min_{X \subseteq  \calC} \cost(q_i,X). \]
The sum of these values over a prefix of the workload is denoted
\[ \textstyle \BASE_n = \sum_{i=1}^n \cost(q_i,S^*_i). \] 
Our analysis makes frequent use of these quantities 
as a simple lower bound on the 
query processing cost that must be paid by any algorithm.
The following result shows that the minimum query cost also
bounds the amount that the work function for an individual state 
increases after each statement.

\hspace{-.265in}\parbox{5in}{ 
\begin{lem}
    \label{lem:wf_slope}
    $\!w_{i+1}(S) \geq w_i(S)\!+\!\cost(q_{i+1},S^*_{i+1})$ for all $i\!\geq\!0$.
\end{lem}
\espace}
\csprf 
For the case $i = 0$, we apply the triangle inequality of $\trans$:
\begin{eqnarray*}
w_1(S) 
 & = &
 \min_{X \subseteq \calC} \{w_0(X) + \cost(q_1,X) + \trans(X,S)\} \\
 & = &
 \min_{X \subseteq \calC} \{\trans(S_0,X) + \cost(q_1,X) + \trans(X,S)\} \\
 & \geq &
 \trans(S_0,S) + \min_{X \subseteq \calC} \cost(q_1,X) \\
 & = &
 w_0(S) + \cost(q_1,S^*_1).
\end{eqnarray*}
For the case $i \geq 1$, note that there exist two index configurations
$P_i,P_{i-1}$ that satisfy the following equations:
\[
\begin{array}{rcl}
 w_{i+1}(S) &=& w_i(P_i) + \cost(q_{i+1}, P_i) + \trans(P_i,S) \\
 w_i(P_i) &=& w_{i-1}(P_{i-1}) + \cost(q_i, P_{i-1}) + \trans(P_{i-1},P_i) 
\end{array}
\]
In other words, the final steps in the path corresponding to $w_{i+1}(S)$ are
$P_{i-1} \rightarrow P_i \rightarrow S$.
The path corresponding to $w_i(S)$ may have a different configuration
as a predecessor to $S$, but we can use the path that passes through
$P_{i-1}$ to bound the value:
\begin{eqnarray*}
w_i(S) 
 &\leq&
 w_{i-1}(P_{i-1}) + \cost(q_i,P_{i-1}) + \trans(P_{i-1},S) \\
 &=&
 w_i(P_i) - \trans(P_{i-1},P_i) + \trans(P_{i-1},S) \\
 &\leq&
 w_i(P_i) + \trans(P_i,S)
\end{eqnarray*}
The triangle inequality yields the third step.
It follows that 
\[
w_{i+1}(S) \geq w_i(S) + \cost(q_{i+1}, P_i) \geq  w_i(S) + \cost(q_{i+1},S^*_{i+1}) \]
as desired.

We next give a result that shows the transition cost of a cyclic sequence
of configurations does not change if we reverse the cycle.

\begin{lem}
    \label{lem:cycle_offset}
    \ Consider the sequence of index configurations
    $S_0,S_1,\dots,S_n,S_0 \subseteq \calC$. The following identity holds:
    \[
    \sum_{i=1}^n \trans(S_{i-1},S_i) + \trans(S_n,S_0)
     = \sum_{i=1}^n \trans(S_{i},S_{i-1}) 
                    + \trans(S_0,S_n)  \]
\end{lem}
\csprf
By induction on $n$. 
The base cases $n=0,1$ are trivial, so we consider
the interesting base case $n=2$. 
For the cycle
$S_0, S_1, S_2, S_0$,
the total transition cost is equal to the sum of
$\trans^+(a)+\trans^-(a)$ over all indices $a$ that occur in
exactly one or two of the sets $S_0,S_1,S_2$. This can be
checked by enumerating the possible sets that such indices
can occur in. By symmetry, the transitions on the reverse cycle
$S_0, S_2, S_1, S_0$
have the same cost.

Now consider the inductive case $n \geq 3$.
By an application of the inductive hypothesis to $S_0,S_1,\dots,S_{n-1},S_0$,

\ospace{\small
\begin{eqnarray*}
   \sum_{i=1}^n \trans(S_{i-1},S_i) + \trans(S_n,S_0) 
   & = &
   \sum_{i=1}^{n-1} \trans(S_{i},S_{i-1}) - \trans(S_{n-1},S_0) \\
   & & {} + \trans(S_0,S_{n-1}) + \trans(S_{n-1},S_n) \\
   & & {} + \trans(S_n,S_0)
\end{eqnarray*}}%
The r.h.s.~contains the cost of the cycle 
$S_0,S_{n-1},S_n,S_0$. Since we assumed $n \geq 3$, we may apply the inductive
hypothesis once more, and replace these terms with the cost of the reverse cycle.
The lemma follows from this substitution.
\cseprf

We can now prove the central result of this section.
The theorem shows a bound on the cost of $\WFA$ that is strictly stronger
than the competitive ratio of Theorem~\ref{thm:wfa_competitive}
when the minimum query costs $\cost(q_i,S^*_i)$ are significant
compared to the cost of the optimal schedule.
The stronger statement is needed to prove Theorem~\ref{thm:wfit_competitive}.

\begin{thm}
\label{thm:wfa_cost}
The total work of $\WFA$ satisfies
\[ 
   \WFA_N - \BASE_N 
   \leq (2^{|\calC|+1}-1) (\OPT_N - \BASE_N) + \alpha \]
where $\alpha$ does not depend on the workload $Q$.
\end{thm}
\csprf
We follow the overall strategy of the proof of
{Lemma~9.3} in~\cite{Borodin:1998ec}, which is an analogous result for
task systems with symmetric transition costs. Our setting differs since
$\trans$ is not symmetric. The original proof also does not
consider the effect of the terms $\cost(q_i,S^*_i)$ that we account for
in the theorem. 

We start with notation borrowed from~\cite{Borodin:1998ec}:
\[\begin{array}{rcl}
\mu &=& \max\{\trans(X,Y) ~|~ X,Y \subseteq \calC\} \\
S_i &=& \mbox{configuration recommended by $\WFA$ for $q_i$} \\
B_i &=& \sum_{S} w_i(S) + \sum_{S \ne S_i} w_i(S)
\end{array}\]
Our eventual goal is to derive separate bounds for 
$\WFA_N$ and $\OPT_N$ with respect to $B_N$, and then combine these bounds.
We first observe that

\ospace{\small
\begin{eqnarray*}
B_{i+1} - B_i
 &\!\!\!=\!\!\!& 
 w_{i+1}(S_i) - w_i(S_{i+1}) 
 + 2 \cdot \hspace{-1em}\sum_{\scriptscriptstyle S \ne S_i,S_{i+1}} 
                 \hspace{-1em} (w_{i+1}(S) - w_i(S)) \\
 & &
 {} + w_{i+1}(S_{i+1}) - w_i(S_{i+1})
    + w_{i+1}(S_{i}) - w_i(S_{i}) \\
 &\!\!\!\geq\!\!\!&
 \rule{0in}{1.5em}
 w_{i+1}(S_i) - w_i(S_{i+1}) \\
 & & 
 {} + (2^{|\calC|+1} - 2)\cdot \cost(q_{i+1},S^*_{i+1})
\end{eqnarray*}}%
by Lemma~\ref{lem:wf_slope}. We can show that
\[ w_{i+1}(S_i) - w_i(S_{i+1}) \geq \trans(S_{i+1},S_i) + \cost(q_{i+1},S_{i+1}) \]
using the same reasoning as~\cite{Borodin:1998ec} (see the original proof
for details%
\footnote{The inequality from~\cite{Borodin:1998ec} reverses the arguments to $\trans$.
In the original proof, the distinction is not important because the transitions
are assumed to be symmetric. In order to prove the competitive ratio in our
setting, the arguments to $\trans$ are reversed in the criteria that $\WFA$
uses to select the next recommendation. 
This results in the slightly different inequality.}).
Hence,
\begin{eqnarray*}
B_{i+1} - B_i 
 &\geq& \trans(S_{i+1},S_i) + \cost(q_{i+1},S_{i+1}) \\
 & &
 {} + (2^{|\calC|+1} - 2)\cdot \cost(q_{i+1},S^*_{i+1}).
\end{eqnarray*}
Summing these inequalities for $0 \leq i \leq N-1$ yields
\begin{eqnarray*}
B_N - B_0 
 &\geq& 
 \sum_{i=1}^N \trans(S_{i},S_{i-1}) + \cost(q_i,S_i) \\
 & &
 {} + (2^{|\calC|+1} - 2)\cdot \BASE_N.
\end{eqnarray*}
The first line of the r.h.s.~is similar to the total work of
$\WFA$ except that the transition costs are reversed. We can
remedy this via Lemma~\ref{lem:cycle_offset}, which leads to
\begin{eqnarray*}
B_N - B_0 
 &\geq& 
 \sum_{i=1}^N \trans(S_{i-1},S_{i}) + \cost(q_i,S_i) \\
 & &
 {} + \trans(S_N,S_0) - \trans(S_0,S_N) \\
 & &
 {} + (2^{|\calC|+1} - 2)\cdot \BASE_N \\
 &=&
 \WFA_N - (2^{|\calC|+1} - 2)\cdot \BASE_N \\
 & &
 {} + \trans(S_N,S_0) - \trans(S_0,S_N)
\end{eqnarray*}
Finally, we can bound $\WFA_N - \BASE_N$ by
\begin{eqnarray*}
\WFA_N - \BASE_N &\leq&
  B_N + (2^{|\calC|+1} - 1) \cdot \BASE_N \\
  & & 
  {} - B_0 - \trans(S_N,S_0) + \trans(S_0,S_N) 
\end{eqnarray*}
To complete the proof, we note that
\[ B_N \leq (2^{|\calC|+1}-1) \cdot \OPT_N + (2^{|\calC|+1}-2)\mu \]
as shown in~\cite{Borodin:1998ec}.
\cseprf

\stitle{Proof of Theorem~\ref{thm:wfa_competitive}}
Rearranging the terms in Theorem~\ref{thm:wfa_cost}, we have
\[
\WFA_N 
 \leq 
 (2^{|\calC|+1}-1)\OPT_N 
 - (2^{|\calC|+1}-2)\BASE_N + \alpha
\]
where $\alpha$ does not depend on the workload $Q$.
Since $(2^{|\calC|+1}-2)$ and $\BASE_N$ are nonnegative,
Theorem~\ref{thm:wfa_competitive} follows.
\cseprf

\stitle{Proof of Theorem~\ref{thm:wfit_competitive}}
We now show the competitive ratio of $\wfaplus$ using a fixed
stable partition $\{C_1,\dots,C_K\}$ of the candidate indices
$\calC$. Our strategy is to use  Theorem~\ref{thm:wfa_cost}
to analyze the recommendations chosen by $\WFA$ within 
each part $C_k$.

We first extend some of our previous notation to describe the
behavior of $\WFA$ in an individual part $C_k$.
We use $\WFA^{(k)}$ to represent an instance of $\WFA$ that
selects recommendations from the part $C_k$ only.
Similarly, $\OPT^{(k)}$ is the idealized algorithm that chooses
the optimal recommendations from $C_k$ with advance knowledge 
of the workload $Q$. Our shorthand for total work
extends naturally, e.g., $\WFA^{(k)}_n$ denotes the total
work of the path that implements the recommendations of $\WFA^{(k)}$
for the first $n$ queries. We denote
\[\begin{array}{rcl}
    S_i 
    &=& \mbox{recommendation of $\wfaplus$ for $q_i$} \\
    S^{(k)}_i \;\equiv\; S_i \cap C_k
    &=& \mbox{recommendation of $\WFA^{(k)}$ for $q_i$} \\
    O_i
    &=& \mbox{recommendation of $\OPT$ for $q_i$} 
\end{array}\]
It follows easily from (\ref{eq:querycost})
that $O_1 \cap C_k, \dots, O_n \cap C_k$
is an optimal path within the part $C_k$ of the stable partition.
A similar fact is:
\[ S^*_i \cap C_k = \arg\min_{X \subseteq C_k} \cost(q_i,X) \]

\fspace\noindent
In other words, the minimum query processing cost using configurations
within $C_k$ is achieved by $S^*_i \cap C_k$.
Thus we use the notation 
\[ \textstyle \BASE^{(k)}_n = \sum_{i=1}^n \cost(q_i,S^*_i \cap C_k) \]
We can also break down the transition cost based on the partition:
\[ \trans(X,Y) = \sum_k \trans(X \cap C_k, Y \cap C_k) \]

\fspace
We need to express the total work of $\wfaplus$ and $\OPT$ 
w.r.t.~the lower bounds $\BASE^{(k)}_N$ in order
to apply Theorem~\ref{thm:wfa_cost}. 
Note that $\OPT$'s recommendations $O_0,\dots,O_N$
obey the following identity:
\begin{eqnarray*}
 & & \cost(q_i,O_i) - \cost(q_i,S^*_i)\\
 & & \ \ \ \ \ = \ \ 
 \benefit_{q_i}(S^*_i,\emptyset) - \benefit_{q_i}(O_i,\emptyset) \\
 & & \ \ \ \ \ = \ \ 
 \sum_k \benefit_{q_i}(S^*_i \cap C_k,\emptyset) 
 - \benefit_{q_i}(O_i \cap C_k,\emptyset) \\
 & & \ \ \ \ \ = \ \ 
 \sum_k \cost(q_i, O_i \cap C_k)
 - \cost(q_i, S^*_i \cap C_k) 
\end{eqnarray*}
Now we rewrite the total work of $\OPT$, offset by $\BASE_N$:

\ospace{\small
\begin{eqnarray*}
 \!\!\!\!&\!\!\!\!& 
 \OPT_N - \BASE_N \\
 \!\!\!\!&\!\!\!\!& \ \ \ \ \ = \ \ 
 \sum_{i=1}^N 
   \trans(O_{i-1},O_i) + \cost(q_i,O_i) - \cost(q_i,S^*_i) \\
 \!\!\!\!&\!\!\!\!& \ \ \ \ \ = \ \ 
 \sum_{i} 
   \trans(O_{i-1},O_i) + 
   \sum_k \cost(q_i,O_i \cap C_k) - \cost(q_i,S^*_i \cap C_k) \\
 \!\!\!\!&\!\!\!\!& \ \ \ \ \ = \ \ 
 \sum_k \sum_i
   \trans(O_{i-1} \cap C_k,O_i \cap C_k) + 
   \cost(q_i,O_i \cap C_k) \\ 
   \!\!\!\!&\!\!\!\!& \hspace{5em} 
   {} - \sum_i \cost(q_i,S^*_i \cap C_k) \\
 \!\!\!\!&\!\!\!\!& \ \ \ \ \ = \ \ 
 \sum_k \OPT^{(k)}_N - \BASE^{(k)}_N
\end{eqnarray*}}%
Applying the same steps to the sequence $S_0,\dots,S_N$, we can derive an analogous
bound for the cost of $\wfaplus$:
\[ \textstyle \wfaplus_N - \BASE_N = \sum_k \WFA^{(k)}_N - \BASE^{(k)}_N. \]
Now by Theorem~\ref{thm:wfa_cost}, we know that 
$\WFA^{(k)}_N - \BASE^{(k)}_N$ is bounded above by
$(2^{|C_k|+1} - 1)(\OPT^{(k)}_N-\BASE^{(k)}_N) + \alpha_k$ where 
$\alpha_k$ does not depend on the workload.
Then we obviously have
\[
\WFA^{(k)}_N - \BASE^{(k)}_N 
 \leq (2^{\cmax+1}-1)(\OPT^{(k)}_N-\BASE^{(k)}_N)
 + \alpha_k
\]
where, as in the statement of Theorem~\ref{thm:wfit_competitive},
$\cmax$ is defined as the maximum of $|C_k|$ for all $k$.
We apply this as follows:

\ospace{\small
\begin{eqnarray*}
 \wfaplus_N 
 \!\!\!\!&=&\!\!\!\!
 \BASE_N + \sum_k \WFA^{(k)}_N - \BASE^{(k)}_N \\
 \!\!\!\!&\leq&\!\!\!\!
 \BASE_N + (2^{\cmax+1} - 1)\sum_k(\OPT^{(k)}_N-\BASE^{(k)}_N) + \sum_k \alpha_k \\
 \!\!\!\!&=&\!\!\!\!
 \BASE_N + (2^{\cmax+1} - 1)(\OPT_N - \BASE_N) + \sum_k \alpha_k \\
 \!\!\!\!&=&\!\!\!\!
 (2^{\cmax+1} - 1)\OPT_N - (2^{\cmax+1} - 2)\BASE_N + \sum_k \alpha_k \\
 \!\!\!\!&\leq&\!\!\!\!
 (2^{\cmax+1} - 1)\OPT_N + \sum_k \alpha_k
\end{eqnarray*}}%
This proves that the competitive ratio of $\wfaplus$ is
$2^{\cmax+1} - 1$.
\cseprf

\section{Proof of Theorem 4.2}
\label{appendix:repartition}

 Theorem~\ref{thm:wfit_partition} states that if $\wfaplus$ uses any fixed stable
partition $\{C_1, \dots, C_K\}$
of the indices $\calC$, it can generate
the same recommendations as the naive application of 
$\WFA$ that jointly tracks all subsets of $\calC$.
Before the main proof, we give one preliminary result that
describes the relationship between the global work function
and the work functions that $\wfaplus$ maintains for each part.

\newcommand{\boldwk}{\boldw^{(k)}}

\begin{lem}
    \label{lem:partitioned_wf}
    Let $\work_n$ be the work function for the workload
    $Q_n$ and the indices $\calC$. Let $\boldwk_n$ be the work 
    function values calculated by $\wfaplus$ for $C_k$ after
    observing $Q_n$. For any $S \subseteq \calC$,
    \[
    \work_n(S) = \sum_k \boldwk_n[S \cap C_k] - (K-1) \sum_{i=1}^n \cost(q_i,\emptyset).
    \]
\end{lem}
\eat{ 
\csprfsketch
The proof is a straightforward induction on $n$. The base case $n=0$
follows by applying the identity
\( \textstyle \trans(X,S) = \sum_k \trans(X \cap C_k, S \cap C_k) \)
to decompose $\work_0(S)$. The inductive case also uses this identity,
the inductive hypothesis, and the fact that
\[\cost(q_n,X)
 =
 \textstyle
 \sum_k \cost(q_n, X \cap C_k) - (K-1) \cost(q_n, \emptyset) \]
for all $X \subseteq \calC$, which is an immediate consequence of  
(\ref{eq:querycost}). The theorem follows by applying these observations
to the definition of $w_n(S)$ and simplifying the result. \cseprf
} 
\csprf
By induction on $n$. For the base case $n=0$, the sum is empty, and
the theorem follows from the identity
\[ \textstyle \trans(X,S) = \sum_k \trans(X \cap C_k, S \cap C_k) \]
mentioned in earlier sections. Specifically, we have
\[\textstyle
  \work_0(S) 
  = \trans(S_0,S) 
  = \sum_k \trans(S_0 \cap C_k, S \cap C_k)
  = \sum_k \boldwk_0(S \cap C_k). \]

To prove the inductive case $n>1$, we observe the following
identity that follows easily from (\ref{eq:querycost}):
for all $X \subseteq \calC$,
\begin{eqnarray*}
\cost(q_n,X)
 &=&
 \textstyle
 \sum_k \cost(q_n, X \cap C_k) - (K-1) \cost(q_n, \emptyset)
\end{eqnarray*}
Recall the definition of the work function
\[ \work_n(S) = 
   \min_{X \subseteq \calC}
   \{\work_{n-1}(X) + \cost(q_n,X) + \trans(X,S) \}
\]
If we apply the inductive hypothesis to $\work_{n-1}(X)$
and also decompose the terms $\cost(q_n,X)$ and $\trans(X,S)$
as shown above, the result simplifies to

\ospace\ospace{\small 
\begin{eqnarray*}
\work_n(S) 
 &=&
 \min_{X \subseteq \calC}
 \textstyle
 \{ \sum_k \boldwk_{n-1}(X \cap C_k) 
           \begin{array}{l} 
               {} \\
                +\ \ \cost(q_n, X\cap C_k) \\
                +\ \ \trans(X \cap C_k, S \cap C_k)\}
               \end{array} \\
 & & {} 
 \textstyle
 - (K-1)\sum_{i=1}^n \cost(q_i,\emptyset).
\end{eqnarray*}}%
The terms in the sum over $k$ each depend on a disjoint part $C_k$, so
the summation can be pulled out of the $\min$ operation to yield

\ospace{\small
\begin{eqnarray*}
\work_n(S) 
 \!\!\!\!&=&\!\!\!\!
 \textstyle
 \sum_k
 \displaystyle
 \min_{X_k \subseteq C_k} 
 \{ \boldwk_{n-1}(X_k) + \cost(q_n, X_k) + \trans(X_k, S \cap C_k) \} \\
 & & {}
 \textstyle \!\!\!\!\!
 - (K-1)\sum_{i=1}^n \cost(q_i,\emptyset) \\
 \!\!\!\!&=&\!\!\!\!
 \textstyle
 \sum_k \boldwk_n[S \cap C_k] - (K-1) \sum_{i=1}^n \cost(q_i, \emptyset).
 \hspace{1em}\csbbox
\end{eqnarray*}}%

In order to prove the equivalence between $\wfaplus$ and $\WFA$,
we must resolve the fact that the selection criteria of $\WFA$ 
are not deterministic: if more 
than one configuration satisfies the criteria, the pseudocode does
not specify which configuration is chosen. Thus, we assume a simple
tie-breaker based on lexicographic ordering, as follows.
Let $\{a_1, \dots, a_{|\calC|}\}$ denote the indices in $\calC$.
If $X,Y \subseteq \calC$ and $X \ne Y$, consider the minimum
value of $d$ where $X$ and $Y$ differ on $a_d$, meaning that
$a_d$ is in the symmetric difference $X \ominus Y$.
The lexicographic tie-breaking rule prefers $X$ to $Y$ iff
$a_d \in X$.

\newcommand{\Sk}{S^k}
\newcommand{\Sl}{S^\ell}
\newcommand{\Shat}{\widehat{S}}

We consider a workload $Q_n$ of length $n$.
Let $S_0$ be the initial configuration,
$S_1, \dots, S_n$ be the recommendations of $\WFA$, and
$\Sk_1, \dots, \Sk_n$ be the recommendations of $\wfaplus$ within
each part $C_k$. The claim of Theorem~\ref{thm:wfit_partition}
can be stated as $S_n = \bigcup_k \Sk_n$ for $n \geq 0$.
The proof proceeds by induction on $n$.
The base case $n=0$ is trivial, as both algorithms start with the same state.

Consider the inductive case $n \geq 1$.
Assume for contradiction that $S_n \ne \bigcup_k \Sk_n$.
Take the minimum $d$ where $a_d \in S_n \ominus \bigcup_k \Sk_n$.
Let $C_\ell$ be the part that contains $a_d$, implying
either $a_d \in \Sl_n - S_n$ or
$a_d \in S_n - \Sl_n$. We first consider the case 
$a_d \in \Sl_n - S_n$. Let $\Shat_n$ denote 
$(S_n - C_\ell) \cup \Sl_n$ which is the result of modifying
$S_n$ to be consistent with $\wfaplus$'s choice within $C_\ell$.
We immediately observe that the lexicographic tie-breaker
prefers $\Shat_n$ to $S_n$. The set 
$\Shat_n$ also satisfies $\WFA$'s explicit tie-breaking constraint
$\Shat_n \in \boldp[\Shat_n]$, by virtue of the fact that both
$S_n$ and $\Sl_n$ satisfy the constraint. Hence, the only possible 
reason that $\WFA$ recommends $S_n$ instead of $\Shat_n$ must be that
$\score(S_n) < \score(\Shat_n)$, i.e.,
\[
\work_n(S_n) + \trans(S_n,S_{n-1}) < \work_n(\Shat_n) + \trans(\Shat_n,S_{n-1}).
\]
We may use Lemma~\ref{lem:partitioned_wf} to decompose both sides of the
inequality according to the stable partition. Since $S_n$ and $\Shat_n$
agree on all indices outside of $C_\ell$, we may cancel terms to yield:

{\small\espace\[
\begin{array}{l}
\boldw_n^{(k)}(S_n \cap C_\ell) + \trans(S_n \cap C_\ell,S_{n-1} \cap C_\ell)
\\ \hspace{20ex} <
\boldw_n^{(k)}(\Sl_n) + \trans(\Sl_n,S_{n-1} \cap C_\ell).
\end{array}
\]}%
I.e., $S_n \cap C_\ell$ has a better score than $\Sl_n$, contradicting the
fact that $\wfaplus$ recommends $\Sl_n$ within $C_\ell$.

The proof when $a_d \in S_n - \Sl_n$ is completely symmetric: we contradict
$\WFA$'s choice of $S_n$ by showing $\Shat_n$ has a lower score.
\cseprf

  \raggedbottom
    
\end{document}